\documentclass[useAMS,usenatbib,usegraphicx]{mn2e}

\title[Empirical test of the halo model]{The distribution of red and
  blue galaxies in groups: an empirical test of the halo model}

\author[A. A. Collister, O. Lahav]{Adrian A. Collister$^{1}$\thanks{aac@ast.cam.ac.uk}, Ofer Lahav$^{2}$
\\ $^{1}$ Institute of Astronomy, University of Cambridge, Cambridge CB3 0HA, UK
\\ $^{2}$ Department of Physics and Astronomy, University College London, Gower Street, London WC1E 6BT, UK}

\date{}

\pagerange{\pageref{firstpage}--\pageref{lastpage}} \pubyear{2004}

\begin{document}

\label{firstpage}

\maketitle

\begin{abstract}
The popular halo model predicts that the
power spectrum of the galaxy fluctuations is
simply the sum of the large scale linear halo-halo power spectrum and the
weighted power spectrum of the halo profile.
Previous studies have derived halo parameters from the observed
galaxy correlation function. Here we test the halo
model directly for self-consistency
with a minimal set of theoretical assumptions
by utilising the 2dF Galaxy Redshift Survey (2dFGRS).
We derive empirically the halo occupation and galaxy radial distributions
in the haloes of the 2dF Percolation-Inferred Galaxy Group (2PIGG)
catalogue.  The mean halo occupation number is
found to be well-fitted by a power-law, $\left<N|M\right> \propto
M^{\beta}$, at
high masses, with $\beta = 1.05, 0.88, 0.99$ for red, blue and all
galaxies respectively (with 1-sigma errors of 15-19\%). We find that
the truncated NFW profile provides a good fit to the galaxy
radial distributions, with concentration parameters
$c=3.9, 1.3, 2.4$  for red,
blue and all galaxies respectively (with  1-sigma errors of
8-15\%).
Adding the observed linear power spectrum to
these results, we compare these empirical predictions of the halo model
with the observed correlation functions for these same 2dF galaxy
populations. We conclude that subject to some fine
tuning it is an acceptable model for the two-point correlations.
Our analysis also explains why the correlation function slope of the red
galaxies is steeper than that of the blue galaxies. It is mainly due to
the number of red and blue galaxies per halo, rather than the
radial distribution within the haloes of the two galaxy species.
\end{abstract}


% Introduction =============================================================

\section{Introduction}
\label{sec.intro}

It is
well known that galaxies of different types are not identically
distributed within galaxy groups (manifested in the well-known phenomenon of morphological
segregation, e.g. \citealt{dressler}; \citealt{postman}; \citealt{goto}). Red, typically
elliptical galaxies, with low
star-formation rates are preferentially found towards the centres of
large groups, while blue, actively star-forming galaxies
dominate in the outskirts of groups and in the field. Furthermore, measurements of the galaxy-galaxy two-point correlation function have
shown that it is well fitted by a power-law over a wide range of
distance scales (e.g. \citealt{peebles}; \citealt{zehavi};
\citealt{hawkins}), and that the slope of
this power-law is a function of galaxy spectral type or colour
(e.g. \citealt{zehavi}; \citealt{norberg_cf};
\citealt{madgwick_cf}): blue galaxies have a shallower two-point correlation
function than the red galaxies. The correlation function of the dark
matter, on the other hand, has a lower amplitude and is far from being a featureless
power-law, thus the galaxy distribution is said to be \emph{biased}
with respect to the dark matter.

Much recent progress towards understanding the nature of galaxy biasing has come through
use of the halo model of large scale structure. The model has its
origins in the work of \citet{neyman_scott}, and was first applied to
continuous density fields by \citet{scherrer}. It has been recently used
in the context of the clustering of both dark matter and
galaxies (e.g. \citealt{peacock}; \citealt{seljak}; \citealt{scoccimarro}; \citealt{berlind_weinberg};
\citealt{maglio}; \citealt*{vdb}). \citet{cooray} provide a detailed
review. In the model, matter in the 
universe is assumed to reside only in discrete haloes. The distribution
of the matter within the haloes gives rise to the non-linear
component of the power spectrum, while the large-scale distribution of the
haloes in space is responsible for the power on linear scales. The
simplicity of the model allows analytic calculations of correlation
functions to be made, so that large parameter spaces can be
investigated at relatively little cost when compared to numerical
simulations of the large scale structure. 

Two ingredients are required to extend the model to galaxy clustering:
the probability distribution for the number of galaxies hosted by a
particular halo, $P(N|M)$, known as the halo occupation distribution (HOD),
and the spatial distribution of galaxies within haloes. The
distribution of the galaxies within haloes is commonly assumed, ad hoc, to be the same as
that of the dark matter; differences in the clustering properties are
then solely due to the choice of halo occupation distribution. We do
not rely on this assumption in our analysis below, but instead obtain the actual galaxy group
radial density distributions observationally. This allows us to
confirm, for observed HODs and radial profiles, the findings of
\citet{sheth_cf}, who used HODs drawn from
semi-analytic models to demonstrate the
dominance of the HOD over the intra-halo distribution in the
correlation function.

Previous studies have assumed the validity of the halo model, and used
the observed galaxy correlation
functions to constrain the halo occupation distribution, or
the radial distribution of galaxies in haloes
(e.g. \citealt{maglio}; \citealt{scranton}; \citealt{vdb}). In
contrast, we take as empirical an
approach as possible. We directly
measure the radial and halo occupation distributions of galaxies in
the 2dFGRS Percolation-Inferred Galaxy Group
(2PIGG) catalogue \citep{ekea}, focusing in particular on the
separate distributions of red and blue galaxies. Having measured these
distributions, no free parameters remain in the model. Since the
correlation functions of these populations
have previously been directly obtained (\citealt{norberg_cf}; \citealt{hawkins}; \citealt{madgwick_cf}), the non-trivial success of the halo model in reproducing the
observed clustering statistics can be directly tested using fully
self-consistent 2dF observations. Requiring the model to account for
the differences in the observed correlation functions of the red and
blue populations is an especially stringent test. 

The structure of this paper is as follows. In Section \ref{sec.formalism}
we describe the halo model in detail. We introduce the 2PIGG catalogue
in Section \ref{sec.2pigg} and describe how the properties of interest are
inferred from the observational data. In Section \ref{sec.hod} the
halo occupation distribution is investigated, and in Section
\ref{sec.rad.dist} we examine the radial distribution of galaxies within the
2PIGGs. Finally, in Section
\ref{sec.cfs}, the halo model is applied to our results, and the
predicted clustering results compared with observations.

% The Halo Model ====================================================================

\section{The halo model formalism}
\label{sec.formalism}

In essence, the halo model exemplifies the natural distinction between linear
(large-scale) and non-linear (small-scale) clustering. Indeed, the two
regimes appear as separate terms in the halo model galaxy
power spectrum:
\begin{equation}
\label{eqn.halo.ps}
  P_\rmn{gal}(k) = P^\rmn{(1h)}_\rmn{gal}(k) + P^\rmn{(2h)}_\rmn{gal}(k),
\end{equation}
with $P^\rmn{(1h)}_\rmn{gal}(k)$ the \emph{intra}-halo, non-linear
term, and $P^\rmn{(2h)}_\rmn{gal}(k)$ the \emph{inter}-halo,
linear term. Although we will later require the two-point correlation
functions, $\xi(r)$, explicit calculation of these involves convolutions
of the halo profiles; we prefer instead to work in Fourier space,
ultimately obtaining the correlation function via the transform
\begin{equation}
\xi_\rmn{gg}(r) = \int \Delta^2_\rmn{gal}(k) \frac{\sin(kr)}{kr}\frac{\rmn{d}k}{k},
\end{equation}
with $\Delta^2_\rmn{gal}(k) = \frac{k^3}{2 \upi^2}
P_\rmn{gal}(k)$, the dimensionless form of the power spectrum.

Within the halo model framework, the components of the galaxy power
spectrum are (e.g. \citealt{cooray})
\begin{equation}
  \label{eqn.1halo}
  P^\rmn{(1h)}_\rmn{gal}(k) = \int \rmn{d}M~n(M)\frac{\left< N(N-1)|M\right>}{\bar{n}_\rmn{gal}^2} |\hat{u}_\rmn{gal}(k|M)|^2,
\end{equation}
and
\begin{equation}
\label{eqn.2halo}
  P^\rmn{(2h)}_\rmn{gal}(k) =
  P^\rmn{(lin)}_\rmn{dm}(k)\left[\int\rmn{d}M n(M) b(M) \frac{\left< N|M\right>}{\bar{n}_\rmn{gal}}\hat{u}_\rmn{gal}(k|M)\right]^2,
\end{equation}
where the integrals are over the halo mass, $M$. In these expressions,
$n(M)$ is the halo mass function, $b(M)$ is the
halo biasing factor, $\left<N|M\right>$ and $\left<
N(N-1)|M\right>$ are the first and second factorial moments of the
halo occupation distribution, $P(N|M)$, respectively,
$\bar{n}_\rmn{gal}$ is the average number density of galaxies, and
$\hat{u}_\rmn{gal}(k|M)$ is the Fourier
transform of the normalised radial distribution of galaxies within
haloes. The following sections describe these terms in more detail.

In keeping with the ethos of this work, we opt to use the simplest
possible implementation of the halo model. We note for completeness
that the implementation may be modified to account for the expectation that the
first galaxy in each halo resides at the halo centre of mass (see
e.g. \citealt{cooray}; \citealt{kravtsov}).

\subsection{Halo mass function}
\label{sec.mass_fn}
The number density of haloes of mass $M$ in space is described by
the mass function, $n(M)$, which we assume to have the form proposed
by \citet{sheth_tormen} (an extension of the \citet{press_schechter} form):
\begin{equation}
  \label{eqn.mf}
n(M)\rmn{d}M = \frac{\bar{\rho}}{M}f(\nu)\rmn{d}\nu,
\end{equation}
\begin{equation}
  \nu f(\nu) = A_\rmn{*}\left(1 + (q\nu)^{-p}\right)\left({q\nu\over 2\upi}\right)^{1/2}\rmn{e}^{-q\nu/2},
\end{equation}
where $\bar{\rho}$ is the background density of the universe, $q=0.707$, $p=0.3$, and normalisation implies
$A_\rmn{*}\approx0.3222$. The mass variable is defined as $\nu\equiv\left(\delta_\rmn{sc}(z)/\sigma(M)\right)^2$,
where $\delta_\rmn{sc}(z)$ is the linear-theory prediction for the
present day overdensity of a region undergoing spherical collapse
at redshift $z$, and $\sigma(M)$ is the r.m.s. variance of the present
day linear power spectrum in a spherical top-hat which contains an average mass
$M$. Note that the mass function depends on the redshift only
through $\delta_\rmn{sc}(z) = \delta_\rmn{sc}(0)/D(z)$, where
$\delta_\rmn{sc}(0) \approx 1.68$, and
$D(z)$ is the linear growth factor, normalized so that $D(0)=1$. The
value of $\delta_\rmn{sc}(0)$ is only weakly sensitive to the cosmological
model \citep*{eke_delta}.

\subsection{Halo biasing}

Haloes are biased tracers of the overall dark matter distribution. 
The degree of bias is a function of the halo mass. Following
\citet{mo_white}, we can write the power spectrum of dark matter
haloes of given masses, $M_1$ and $M_2$ as 
\begin{equation}
  P_\rmn{hh}(k; M_1, M_2) = b(M_1)~b(M_2)~P_\rmn{dm}(k).
\end{equation}
We further assume $P_\rmn{dm}(k) = P^\rmn{(lin)}_\rmn{dm}(k)$, the
linear dark matter power spectrum, since inter-halo correlations
are only important on large, quasi-linear scales. \citet{sheth_tormen}
derive the required halo bias factors,
\begin{equation}
  \label{eqn.halo_bias}
  b(M) = 1 + \frac{q\nu-1}{\delta_{sc}(z)} + \frac{2p/\delta_{sc}(z)}{1 + (q\nu)^p}.
\end{equation}
with $p$ and $q$ taking the values given in Section \ref{sec.mass_fn}.

\subsection{Galaxy distribution}

The halo occupation distribution, $P(N|M)$, appears in the power
spectrum (equations \ref{eqn.1halo} and \ref{eqn.2halo}) through its
first and second factorial moments, $\langle N|M \rangle$ and $\langle
N(N-1)|M \rangle$ respectively. The galaxy number density is given by
\begin{equation}
  \label{eqn.ndens}
  \bar{n}_\rmn{gal} = \int \left< N|M\right>n(M)~\rmn{d}M.
\end{equation}

The spatial distribution of galaxies within haloes is assumed to be spherically
symmetric about the halo centre, so that the density profile,
$\rho_\rmn{gal}(r|M)$, is a function of $r$ only for a halo of a
given mass $M$. The profile is normalized so that
$\int_0^{r_\rmn{vir}} \hat{\rho}_\rmn{gal}(r|M)~4 \upi r^2\rmn{d}r
= 1$ ($r_\rmn{vir}$ will be defined in equation \ref{eqn.rvir}), and the
Fourier transform of the normalized profile is denoted by
$\hat{u}_\rmn{gal}(k|M)$. 

We attempt to directly measure the radial density and halo occupation distributions
in subsequent sections.

\subsection{Galaxy bias}

On distance scales for which the inter-halo term is important,
$u_\rmn{gal}(k|M) \approx 1$ is a good approximation. The integral
on the right-hand side of equation \ref{eqn.2halo} is then independent of
scale and we can
re-write the relations as:
\begin{equation}
  \label{eqn.bias_param}
  P^\rmn{(2h)}_\rmn{gal}(k)
  = b^2_\rmn{gal} P^\rmn{(lin)}_\rmn{dm}(k),
\end{equation}
where we have defined the galaxy bias parameter as:
\begin{equation}
  \label{eqn.gal_bias}
  b_\rmn{gal} \equiv \int \rmn{d}M~n(M)~b(M) \frac{\left<N|M\right>}{\bar{n}_\rmn{gal}}.
\end{equation}

\subsection{An empirical approach to the halo model}

As we have stressed, our aim is to rely on observations wherever
possible, avoiding model-dependent assumptions.
Our underlying assumption is that the observed galaxy groups represent
the haloes. Even if this assumption is not perfect it is very likely that
there is a simple ranking relation between the observed galaxy groups and the
underlying dark matter haloes. Although we could take for $n(M)$ the histogram of the
observed groups versus their estimated mass, we prefer to use the
more robust mass function $n(M)$ given by equation \ref{eqn.mf}.
Future group samples with accurate masses will allow us to use
them directly for $n(M)$.

Equation \ref{eqn.1halo} requires knowledge of the occupation
quantities $\left<N|M\right>$, $\left<N(N-1)|M\right>$
and the radial profile $\rho_\rmn{gal}(r)$ which we shall determine
directly from the 2PIGG sample per galaxy type.
The mean number of galaxies (equation \ref{eqn.ndens}) then follows from the above
$n(M)$ and $\left<N|M\right>$.

The halo-halo correlation power spectrum
$P^\rmn{(2h)}_\rmn{gal}(k)$ could in principle be taken directly from
the group-group power spectrum.
The group-group correlation functions have actually been derived
by \citet{padilla} and \citet{yangb}.
However, biases in mapping the groups to haloes may make this approach
somewhat inaccurate with the present data.
While this should be possible with future group samples
here we follow the approximation given by equations \ref{eqn.bias_param} and \ref{eqn.gal_bias}.
In fact, as shown later in Section \ref{sec.bias}, we find that $b_\rmn{gal}$ is
close to unity, interestingly in accord with the biasing derived
from the 2dF linear galaxy-galaxy
power spectrum \citep{percival} combined with pre-WMAP
CMB measurements \citep{lahav}.
In practice we use equations \ref{eqn.bias_param} and \ref{eqn.gal_bias}
with the linear power spectrum of the dark matter,
$P^\rmn{(lin)}_\rmn{dm}(k)$, which has been well constrained by
observations of the Cosmic Microwave Background (CMB).
The galaxy power
spectrum on linear scales has also been measured observationally,
for example by \citet{percival} from the 2dFGRS, and \citet{tegmark}
from the Sloan Digital Sky Survey. The shapes of these power spectra
have been shown to be consistent with a flat universe $\Lambda$-CDM
matter power spectrum, with present epoch $\Omega_\rmn{m}\sim0.3$ (\citealt{percival}; \citealt{efstathiou}; \citealt{cole}). We therefore assume this
form for $P^\rmn{(lin)}_\rmn{dm}(k)$, adopting the WMAP normalisation
$\sigma_{8}=0.9$ \citep{spergel}  and the biasing parameter derived
in Section \ref{sec.bias}.

We emphasize again that although we have to compromise here by making
several theoretical assumptions, it would be possible in the future
to use galaxy and group catalogues to test the halo model almost without
any theoretical prior.

% The Data ==================================================================

\section{Galaxy group catalogue}
\label{sec.2pigg}

We assume that galaxy groups are representative of the underlying dark
matter haloes. The 2dFGRS Percolation-Inferred Galaxy Group catalogue (2PIGG;
\citealt{ekea}) is currently the largest homogeneous sample
of galaxy groups publicly available. It comprises $\sim$29,000 groups containing
at least two galaxy members, which host a total of $\sim$105,000
galaxies. As described in the following subsections, we apply a number of cuts
to the catalogue in order to improve the quality of the
sample used in our analyses (our final sample comprises $\sim 3000$ groups).

An independent attempt at constructing a galaxy group catalogue for
the 2dFGRS has been made by \citet{yanga}, using a halo model-based algorithm. Although it is not the aim
of this work to perform a comparative study of the two
catalogues we note relevant differences where appropriate.

The 2PIGG groups were identified from the 2dFGRS by means of a
friends-of-friends (FOF) percolation
algorithm. \citet{ekea} tested the algorithm on mock samples generated from cosmological
dark matter simulations in order to optimize the mapping between the
recovered galaxy groups and the actual dark matter haloes;
this strengthens the case for our prior assumption that galaxy groups
may be identified with the true bound structures in the dark matter distribution.

\citet{ekea} tuned their group-finding algorithm so as to maximize the
completeness of the recovered groups. Consequently, very few true
group members are erroneously excluded, but this is at the expense of increased
contamination by interlopers (i.e. field galaxies assigned to
groups, or false groups comprised entirely of field galaxies). The
severity of the contamination increases with redshift; following
\citet{ekeb}, we discard groups at redshifts greater than $z=0.12$. At
this redshift the total number of field galaxies included in groups
rises to $\sim50$ per cent of the total number
of true group members, and the number of
interlopers increases very rapidly with redshift beyond this
point. This cut leaves $\sim$16,000 remaining groups. We note that the
halo-based group finder of \citet{yanga} achieves similar
completeness to standard FOF but reduces
the contamination level, typically by a factor of two (for group masses
$M \ga 10^{14}h^{-1}\rmn{M}_{\sun}$).

\subsection{Spectral classification}
\label{sec.eta}
\citet{madgwick_eta} used a principal component analysis of the 2dFGRS dataset
to define, $\eta$, a continuous parametrization of spectral type. This
parameter is most strongly correlated with the current star formation
rate in each galaxy, but is also a good indicator for
morphological type and colour. Following \citet{madgwick_cf}, we
broadly classify galaxies in our sample using a cut at $\eta =
-1.4$. We label galaxies with $\eta > -1.4$ (relatively active) as
\emph{blue} galaxies, and those with $\eta < -1.4$ (relatively
passive) as \emph{red}. We exclude from
our analysis any group which does not have a measurement of $\eta$ for
\emph{all} its member galaxies. This reduces the sample to $\sim$14,000 groups.

\subsection{Mass estimation}

The majority of the 2PIGGs have a measurement for the one-dimensional velocity
dispersion, $\sigma_\rmn{v}$. This can be used to estimate the group mass
as
\begin{equation}
  \label{eqn.2pigg_mass}
  M = A \frac{\sigma_\rmn{v}^2 R_\rmn{rms}}{G},
\end{equation}
where $R_\rmn{rms}$ is the r.m.s. projected separation from the central galaxy of
the remaining galaxies assigned to the group. Through use of
simulated mock surveys, \citet{ekea} derive the value $A=5.0$ by
requiring that the estimated mass be unbiased with respect to
that of the underlying dark matter haloes. The simulated dark
matter haloes are identified using a friends-of-friends algorithm,
and the recovered haloes have mean spherical overdensities of $\sim
200$ times the background density (Eke, private communication). \citet{sheth_tormen} define the mass
of a halo to be that enclosed within such an overdensity, thus it is
valid to identify the mass estimate of equation \ref{eqn.2pigg_mass} with
the halo mass used throughout Section \ref{sec.formalism}. 

The measurement of the velocity dispersion is very unreliable for
groups with a small number of observed
galaxies,
and even for groups with large memberships we must be
wary of the significant scatter in the relation between group mass and the velocity
dispersion (see fig. 3 of \citealt{ekeb}). \citet{yanga} have independently constructed a galaxy group catalogue for
the 2dFGRS. They investigate
the reliability of dynamical mass estimates, and propose an
alternative halo mass assignment based on total group luminosity. This
gives them a particular advantage at low group masses where the
dynamical masses are especially unreliable.

To avoid the poorest mass estimates we retain only those groups with
at least four observed galaxy members. \emph{The final sample contains
3,147 groups hosting a total of 25,118 galaxies (of which 12,851 are
blue and 12,267 are red)}.

In order to maintain internal consistency regarding the definition of the group
mass, we
define the group virial radius,
\begin{equation}
\label{eqn.rvir}
r_\rmn{vir} = \left(\frac{3 M}{4\upi \Delta \bar{\rho}}\right)^\frac{1}{3},
\end{equation}
to be that enclosing a spherical overdensity $\Delta = 200$
times the background density of the universe (at the group redshift).

\subsection{Group membership}
\label{sec.grp.mem}

In order to estimate the true galaxy membership of the 2PIGGs it
is necessary to correct for (i) the flux limit, and (ii) the
incompleteness of the 2dF survey due to e.g. constraints on fibre
positioning. \citet{ekea} compute for each galaxy a weight, $w_j \geq
1$, to account for the local incompleteness of the survey. These weights only account for missed galaxies which are brighter than the local
magnitude limit, $b_\rmn{J,lim}$; the flux limit is not yet accounted
for.

In order for the group membership to be well-defined observationally,
a limiting absolute magnitude, $M_{b_\rmn{J},\rmn{com}}$, must be specified. We explain in Section
\ref{sec.cfs} how our choices for this limit are dictated by the samples used in determining the observed correlation
functions; the values we use are $M_{b_\rmn{J},\rmn{com}} -
5\log_\rmn{10}h = -19.25, -19.04, -19.50$ for red, blue and all
galaxies respectively.

The faintest detectable absolute magnitude at
distance $x$ is given by $M_{b_\rmn{J}}^{(i)}(x) = b_{\rmn{J,lim}} - 25 - 5\log(x) -
K^{(i)}(x)$ for galaxies of spectral type $i$, where $K^{(i)}(x)$ is
the $K$-correction (measured for each spectral type by
\citealt{madgwick_eta}). Of all the galaxies more luminous than the
threshold $M_{b_\rmn{J},\rmn{com}}$, the fraction which are detectable
at distance $x$ is given by the \emph{selection function}:
\begin{equation}
\label{eqn.sel_fn}
\varphi^{(i)}(x) = \frac{\int_{-\infty}^{M_{b_\rmn{J}}^{(i)}(x)} \phi^{(i)}(M)~\rmn{d}M}{\int_{-\infty}^{M_\rmn{max}} \phi^{(i)}(M)~\rmn{d}M},
\end{equation}
where $M_\rmn{max} =
\rmn{max}(M_{b_\rmn{J}}^{(i)}(x),M_{b_\rmn{J},\rmn{com}})$.
A further complication is that galaxies with $b_\rmn{J} < 14$ were removed from the
redshift catalogue; the lower integration limit of the numerator in equation
\ref{eqn.sel_fn} is adjusted to account for this.

Since we are dealing with galaxies in groups we choose to
use luminosity functions (LFs)
specific to the populations of these over-dense regions, rather
than those of the whole 2dF sample. \citet{croton} have derived 2dF LFs as a function of density environment and per
spectral type. We use their `cluster' LFs for red and blue galaxies to compute the selection
function for the 2PIGGs.

We can now assign to each galaxy a weight $w'_j$, which accounts for both
the local incompleteness and the luminosity and flux limits. This
weight is defined as
\begin{equation}
\label{eqn.weights}
w'_j = \left\{ \begin{array}{ll}
    w_j/\varphi^{(i)}(x) & (M_{b_\rmn{J}} < M_{b_\rmn{J},\rmn{com}}) \\ 
    0 & (M_{b_\rmn{J}} > M_{b_\rmn{J},\rmn{com}})
  \end{array} \right.
\end{equation}
for a galaxy of spectral type $i$ hosted by a group at distance $x$, where $M_{b_\rmn{J}}$ is the absolute magnitude of the galaxy in
question. The galaxy membership of a group, complete to the absolute
magnitude limit, may finally be obtained by summing these weights over
the galaxies assigned to the group.

% Halo Occupation Distribution ===========================================================

\section{Halo occupation distribution}
\label{sec.hod}

\begin{figure}
  \includegraphics[width = 0.5\textwidth]{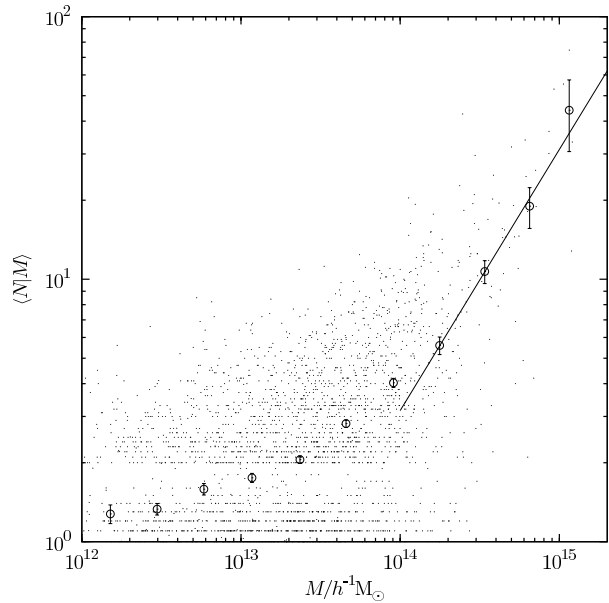}
  \caption{\label{fig.hod.all} Total galaxy occupancy of 2PIGGs
  (above the minimum absolute magnitude limit $M_{b_\rmn{J},\rmn{com}} - 5 \log_\rmn{10}h = -19.50$). Points
  represent the individual groups used in the analysis. Open
  circles show the average occupancy in bins of the group mass, with
  1-sigma statistical errors. The
  dashed line shows the best-fitting power-law (see
  Table \ref{tab.hod}). The fitting was performed only over the range
  $M > 10^{14}h^{-1}\rmn{M}_{\sun}$ since our exclusion of groups
  with fewer than four observed members means that we expect to
  systematically overestimate the halo occupation for less massive groups.}
\end{figure}

\begin{figure}
  \includegraphics[width = 0.5\textwidth]{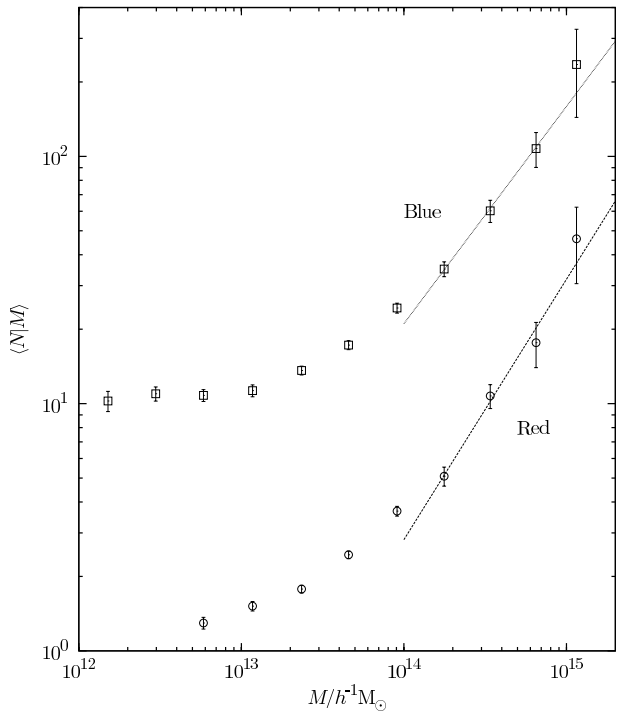}
  \caption{\label{fig.hod.spec} As Fig. \ref{fig.hod.all}, but
  showing separately the mean occupation numbers for blue
  and red galaxies (for clarity, the results for blue galaxies have
  been translated up one decade, and individual groups are not plotted).}
\end{figure}

Having estimated the group membership as described in Section
\ref{sec.grp.mem}, we calculate the mean halo occupation numbers as a
function of group mass (Figs \ref{fig.hod.all} and
\ref{fig.hod.spec}). The error bars on these
points are purely statistical: the uncertainty in the mass estimates is
not directly accounted for. \citet{berlind_weinberg} have used
simulated galaxy group data to gauge the impact of this dispersion on the
determination of $\langle N|M\rangle$. Amongst their conclusions they
find:

(i) The exclusion of groups with fewer than four
observed members is responsible for $\langle N | M \rangle$ being
systematically overestimated at low masses
($M\la10^{14}h^{-1}\rmn{M}_{\sun}$).\footnote{Note that, after the
  imposition of the absolute magnitude limits, $M_{b_\rmn{J},\rmn{com}}$, our sample does in fact include groups with fewer
  than four members.}

(ii) At higher masses, the measured amplitude is likely to be biased
relative to the true value. This is due to the decrease in group abundance
with increasing mass: of the groups scattered into a particular mass
range, the majority originate from lower masses (and therefore have a
lower average occupation number).

(iii) Despite the scatter, the recovered slope at
high group masses is not significantly biased (for sufficiently steep $\langle N
| M \rangle$; see below). 

In the particular case of the 2PIGG catalogue, the impact of (ii) will be countered by the relatively
high number of interlopers (these competing effects are of a roughly similar
magnitude). Furthermore, we note that equations \ref{eqn.1halo} and
\ref{eqn.2halo} do not directly depend on the amplitude of
  $\langle N | M \rangle$ since $\bar{n}_\rmn{gal}$ scales identically,
so the expected bias in our measurement is
not a great concern in this context (but see Section \ref{sec.lowmass}).

Semi-analytic models (e.g. \citealt{benson}; \citealt{diaferio}) unanimously predict that
$\langle N | M \rangle$ should tend to a power-law at high
masses. The behaviour at low masses is expected to be somewhat more complicated,
particularly when the occupation numbers of red and blue
sub-populations are considered separately. Unfortunately, as \citet{berlind_weinberg}
predict, we are unable to reliably probe the halo occupation distribution
for masses less than $\sim10^{14}h^{-1}\rmn{M}_{\sun}$. In the
interests of keeping our results as empirical as possible, we
therefore adopt
the simplest possible model for the mean occupation number: a single
power-law with a cut-off at low mass. Thus,
\begin{equation}
  \label{eqn.hod}
  \langle N | M \rangle = \left\{ \begin{array}{ll}
    \left(M/M_0\right)^\beta & (M \ge M_\rmn{cut}) \\
    0 & (M < M_\rmn{cut}) 
  \end{array} \right.
\end{equation}
where $M_0$ and $\beta$ are free parameters. \citet{berlind_weinberg} suggest that
$\beta$ may be reliably recovered, despite the dispersion in the mass,
for $\beta \ga 0.7$.

Using only the data for groups with  $M >
10^{14}h^{-1}\rmn{M}_{\sun}$, we have fitted power-laws to the measured mean occupation numbers;
the best-fitting parameters are given in Table \ref{tab.hod} and the
fits are overlaid on the results in Figs \ref{fig.hod.all} and
\ref{fig.hod.spec}. In this high mass limit the power-law provides a
good fit. We find $\beta > 0.7$ for all three samples, so these
estimates of the high-mass slope ought not to be seriously affected by the
scatter in the mass. 

Our results for the power-law slope at high group masses are in good
agreement with previous work. For example,
\citet{diaferio} make simple fits to the semi-analytic models of
\citet{kauffmann} and find power-laws with $\beta_\rmn{red} = 0.9$ and
$\beta_\rmn{blue} = 0.8$ (but note that their definitions of `red' and
`blue' differ from ours). \citet{maglio} use the observed 2dF correlation
functions to constrain the HOD. They use the same cut on $\eta$
(see Section \ref{sec.eta}) to define red and blue populations, and find for their best-fitting models
$\beta_\rmn{red} = 1.1^{+0.1}_{-0.2}$ and $\beta_\rmn{blue} =
0.7^{+0.2}_{-0.1}$ in the high-mass limit.

\begin{table}
\caption{Best-fitting parameters to $\langle N | M \rangle =
  (M/M_0)^\beta$. Errors are 1-sigma, marginalized over the
  other fitted parameter in each case.}
\label{tab.hod}

\centering
\begin{tabular}{lcc}

\hline
& $\log(M_0/h^{-1}\rmn{M}_{\sun})$ & $\beta$
\\
\hline
All galaxies & $ 13.50^{+0.13}_{-0.19}$ & $ 0.99^{+0.15}_{-0.17}$
\\
Red & $ 13.57^{+0.13}_{-0.18}$ & $ 1.05^{+0.17}_{-0.19}$
\\
Blue & $ 13.63^{+0.13}_{-0.18}$ & $ 0.88^{+0.16}_{-0.17}$
\\
\hline
\end{tabular}
\centering
\end{table}

\subsection{Second moment}

The second moment $\left<N(N-1)|M\right>$ of $P(N|M)$ is usually
expressed in terms of the parameter $\alpha(M)$, where
\begin{equation}
\label{eqn.alpha}
\left<N(N-1)|M\right> = \alpha^2(M) \left<N|M\right>^2.
\end{equation}
A pure Poisson distribution has $\alpha(M) = 1$, but semi-analytic models
suggest that in fact $P(N|M)$ has a sub-Poisson distribution,
i.e. $\alpha(M) < 1$, particularly at low halo
masses. \citet{kravtsov} and \citet{casas-miranda} discuss models for
the sub-Poissonian behaviour of the HOD.

\begin{figure}
  \includegraphics[width = 0.5\textwidth]{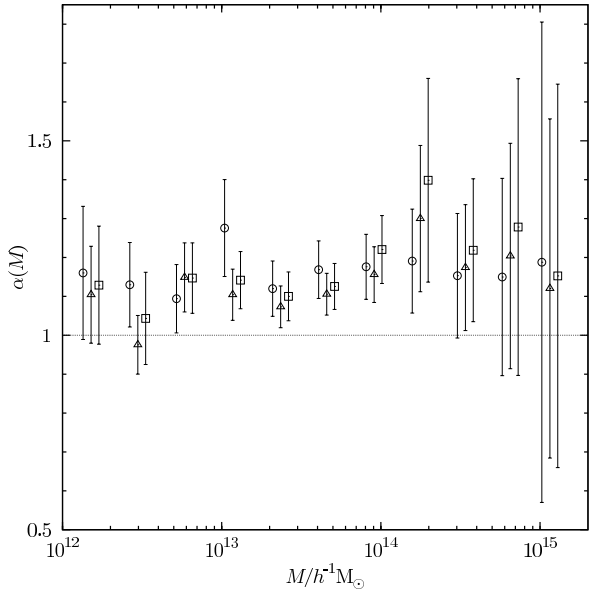}
  \caption{\label{fig.alpha} Measured values of $\alpha(M)$ (defined in
  equation \ref{eqn.alpha}), for blue (circles), red (squares), and all
  galaxies (triangles). For clarity, points for blue and red
  galaxies are displaced slightly either side of the true bin
  mass.}
\end{figure}

We directly measure $\left<N(N-1)|M\right>$ from the 2PIGG data and
show the results in Fig. \ref{fig.alpha} (in the form of
$\alpha$). Contrary to expectations we find $\alpha(M) > 1$  across
the entire mass range for all galaxy types: averaged over all
masses, $\alpha = 1.17 \pm 0.06, 1.16 \pm 0.06, 1.13\pm0.05$ for red,
blue and all galaxies respectively. This is undoubtedly an effect
of the dispersion in the mass estimates: $\left<N(N-1)|M\right>$ describes the
width of $P(N|M)$, but the large scatter in the mass
leads to a broadening of the observed distribution and hence an
overestimate of $\left<N(N-1)|M\right>$. It is especially difficult to
obtain masses for low-mass, low-occupancy groups, and it is therefore
neither surprising nor significant that we do not observe the expected sub-Poissonian
behaviour in this regime. Our measured $\alpha(M)$ can thus only be
considered as an upper limit. Guided by our observations, and in the
interests of simplicity we assume $\alpha(M) = 1$ for all halo masses
when calculating correlation functions in Section \ref{sec.cfs}. We
note that \citet{yangc}, using their alternative mass determination, do
find the expected sub-Poissonian behaviour for low-mass groups.

\subsection{Extrapolation to low mass}
\label{sec.lowmass}

As we mention above, observational difficulties mean we are unable to
recover the halo occupation distribution for haloes less massive
than around $10^{14}h^{-1}\rmn{M}_{\sun}$. In particular, we cannot directly
determine the minimum halo mass, $M_\rmn{cut}$, which may host a
galaxy. However, under the assumption that the HOD is well described
by equation \ref{eqn.hod}, the one remaining free parameter,
$M_\rmn{cut}$, can be constrained by requiring that we recover the
correct average space density of galaxies.

The actual (observed) galaxy number density may be obtained directly from the
luminosity function (integrated over the range of luminosity required); for this purpose we use the
2dF luminosity functions per type (regardless of environment) of
\citet{madgwick_eta}. $M_\rmn{cut}$ can then be constrained by
requiring that equation \ref{eqn.ndens} agrees with this observed value.

\begin{table}
\caption{Galaxy number densities (for galaxies brighter than the
  absolute magnitude limits derived in Section \ref{sec.cfs}) computed from the
  luminosity functions of \citet{madgwick_eta}. The quoted 1-sigma errors are estimated from the
  uncertainties in the parameters of the luminosity functions.}
\label{tab.mcut}

\centering
\begin{tabular}{lc}

\hline
& $\bar{n}_\rmn{g}~(10^{-3} h^3 \rmn{Mpc}^{-3})$
\\
\hline
All galaxies & $5.06 \pm 0.49$
\\
Red & $3.90 \pm 0.35$
\\
Blue & $4.80\pm0.27$
\\
\hline
\end{tabular}
\centering
\end{table}

The values we obtain for $M_\rmn{cut}$ through this method are unphysically
low (in fact, for the combined and red samples it is not possible to
match the observed number density even by allowing $M_\rmn{cut} =
0$). This might be attributed in part to the underestimation of the
amplitude of $\langle N | M \rangle$ predicted by
\citet{berlind_weinberg}, but is most likely due to the overly
simplistic HOD we have adopted (equation \ref{eqn.hod}). It is more
usual to have $\langle N | M \rangle$ tend asymptotically to unity
at low mass (e.g. \citealt{kravtsov}), significantly increasing the number of galaxies
hosted by haloes in this mass range relative to the pure power-law we
have adopted. We consider the impact of these issues on the halo model
correlation functions in Section \ref{sec.sensitive}. For the purposes
of Section \ref{sec.cfs} we adopt $M_\rmn{cut}=0$ when using the
simple power-law HOD (equation \ref{eqn.hod}).

% Radial Distribution ====================================================

\section{Radial distribution of galaxies in groups.}
\label{sec.rad.dist}

We wish to obtain the group galaxy density profile given the group
mass: $\rho_\rmn{gal}(r|M)$. Unfortunately the virial motions of
galaxies in groups give rise to redshift distortions
(e.g. the so-called fingers-of-god), so that redshift information alone is not
suitable for determining the three-dimensional galaxy distribution in
groups. We can therefore only measure the surface density projected
along the line-of-sight, $\Sigma_\rmn{gal}(R)$ (where the upper-case $R$ denotes
\emph{projected} distance from the group centre). Only if
spherical symmetry is assumed can we unambiguously infer
$\rho_\rmn{gal}(r|M)$ from $\Sigma_\rmn{gal}(R)$.

\citet{ekea} systematically label one member of each 2PIGG as the
central galaxy. For each group we calculate the
projected angular separation from the central galaxy
of the remaining group members. Unfortunately, even for the richest
groups, there are insufficent galaxies to allow the radial
distribution to be recovered with reasonable signal-to-noise for an
individual group. We
therefore scale the projected separations in each group by its
virial radius to allow like-for-like stacking of
groups over a range of masses. The total surface density of galaxies is then
calculated by summing the weights $w'_j$ in radial bins. Summing the
weights ensures that the profile is correctly
normalized; although the actual value of the amplitude is of no consequence
in the context of the halo model, it is interesting to compare the relative
number densities of red and blue galaxies as a function of radius. We
find that the shape of the profile is not altered if we use a simple number count
instead of summing the weights. Futhermore, the
shape of the profile is also found to be insensitive to the limiting absolute magnitude
used and we therefore choose to use a considerably fainter limit,
$M_{b_\rmn{J},\rmn{com}} - 5 \log_\rmn{10}h = -17.5$, than used to
determine the group membership, so as to increase the number of
galaxies admitted and hence improve the statistics.

The design of the 2dF spectrograph means that fibres cannot be positioned
within $\sim30\arcsec$ of one another, hence there is effectively an
exclusion zone of this angular radius around the central galaxy in each
group. Furthermore, such close-pair
incompleteness also exists in the 2dF's parent APM catalogue;
\citet{vdb_radial} show that these have an equally
important impact on the measured surface density at small
separations. The severity of the problem is a decreasing function of
the apparent angular size of a group. We have therefore imposed an additional selection on
the sample used in this section, discarding all groups having a virial angular
diameter less than $600\arcsec$. Since fibre collisions occur on
angular scales $\la30\arcsec$, this confines the affected region to
projected separations from the group centre of
$R\la0.05~r_\rmn{vir}$. This choice of cut-off optimizes the
balance between the increased statistical noise resulting from using
fewer groups, and the improved constraints on the profile shape achieved by probing nearer
to the centre. The remaining sample comprises 1,619 groups containing
a total of 16,749 galaxies.

\begin{figure}
  \includegraphics[width = 0.5\textwidth]{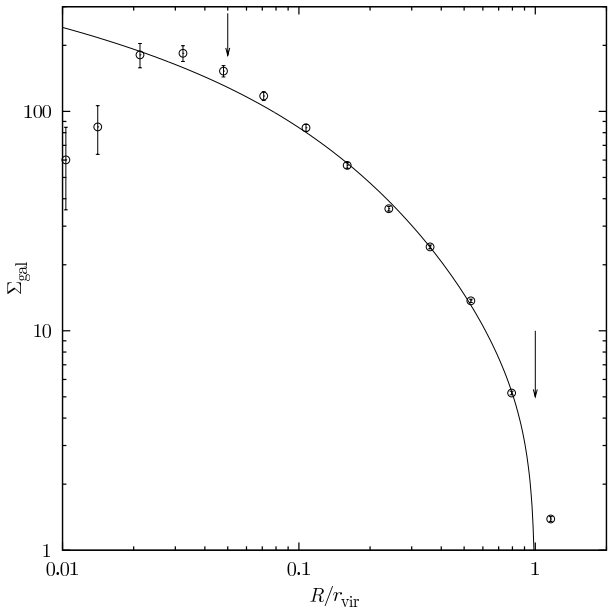}
  \caption{\label{fig.allgals.prof} Projected galaxy density profile
  from stacking all groups in our sample, scaled by their
  respective virial radii. The best-fitting NFW profile
  ($c=3.4$) is overlaid. The decline in
  surface density within $R\la0.05~r_\rmn{vir}$ is due to 
  close-pair incompleteness; the angular diameter selection described in the
  text ensures that its influence is confined to this region for all the groups included in
  this composite profile. Arrows indicate the radial range over
  which the profile fitting was performed.}
\end{figure}

\begin{figure}
  \includegraphics[width = 0.5\textwidth]{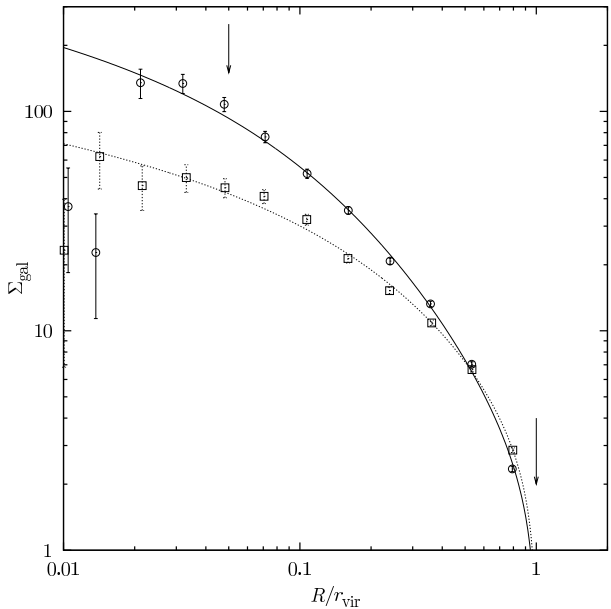}
  \caption{\label{fig.bytype.prof} As Fig. \ref{fig.allgals.prof}
    but showing separately the results for red (circles) and
    blue (squares) galaxies. Red galaxies have a more
    concentrated profile than blue galaxies, and
    dominate the number counts towards the group centre.}
\end{figure}

The observed surface density profiles are shown in Fig.
\ref{fig.allgals.prof} for all galaxy types, and in Fig. \ref{fig.bytype.prof} for
blue and red galaxies separately. The decline in surface density due
to close-pair incompleteness in the region $R\la0.05~r_\rmn{vir}$ is
very apparent. We note also that the surface density in not uniformly
zero outside $r_\rmn{vir}$, as it ideally ought to be since the
group-finding algorithm only admits galaxies which are deemed to be
bound to a group (i.e. they lie within that group's virial
radius). This is due mainly to the scatter in the mass estimates,
which translates into a scatter in $r_\rmn{vir}$ so that there is
uncertainty in the radial scaling of the groups. We assess the impact
of this effect in Appendix \ref{sec.scaling}.

\subsection{Profile fitting}
\label{sec.prof.fit}

Obtaining an analytic fit to the observed profiles will greatly
facilitate their use in the halo model. Simulations suggest that dark
matter halos obey a universal density profile of the form \citep*{nfw}
\begin{equation}
  \label{eqn.nfw}
  \rho(r) = \frac{\rho_{s}}{r/r_\rmn{s}\left(1+r/r_\rmn{s}\right)^2}
\end{equation}
where $r_\rmn{s}$ is a characteristic scale radius, and
$\rho_\rmn{s}$ sets the amplitude of the profile. This NFW profile
is often alternatively expressed in terms of the concentration
parameter, $c = r_\rmn{vir}/r_\rmn{s}$. While there is no physical
prerequisite that galaxies should follow the same profile as the dark
matter, similar previous studies (e.g. \citealt*{lin}) have found the
NFW profile to provide a good fit to the radial distribution of
galaxies, and it therefore constitues the most promising initial guess
for the profile.

In order to attempt fitting of the NFW profile to our observations we require its
plane projection,
\begin{equation}
\label{eqn.projection}
\Sigma(R) = 2 \int^{r_\rmn{max}}_{R} \rho(r) \frac{r~\rmn{d}r}{\sqrt{r^2 - R^2}}.
\end{equation}
In the case $r_\rmn{max} = \infty$, equation \ref{eqn.projection} has an analytic
solution (e.g. \citealt{bartelmann}),  but we require
$r_\rmn{max} = r_\rmn{vir}$ since we assume the profile to be
truncated at the virial radius. We therefore derive the
projected surface density profile numerically from equation
\ref{eqn.projection}, with the integration region bounded at the virial radius.

\citet{nagai} discuss the problem of degeneracy between
the concentration parameter and amplitude, $\rho_\rmn{s}$, of the NFW profile. They
suggest that the amplitude be re-expressed in terms of the 
total number of galaxies contained within the virial radius:
\begin{equation}
N_\rmn{vir} = \int^{r_\rmn{vir}}_{0} \rho(r)~4\pi r^2
\rmn{d}r = 4\pi \rho_\rmn{s} r_\rmn{vir}^3\frac{g(c)}{c^3}
\end{equation}
where $g(c) = \ln(1+c) - c/(1+c)$.
\citet{nagai} were interested in fitting to simulated data for which
$N_\rmn{vir}$ could be trivially obtained, reducing the problem to a
one-parameter fit. The peculiar velocity distortions in the 2dF redshifts
make it impossible to determine $N_\rmn{vir}$ reliably for the 2PIGGs,
hence we are forced to allow the amplitude of the profile to be a
second free parameter. We prefer to use this form none the less, since
$c$ is far less degenerate with $N_\rmn{vir}$ than it is with
$\rho_\rmn{s}$, and we find $N_\rmn{vir}$ a more intuitive parametrization
of the amplitude.

\begin{figure}
  \includegraphics[width = 0.5\textwidth]{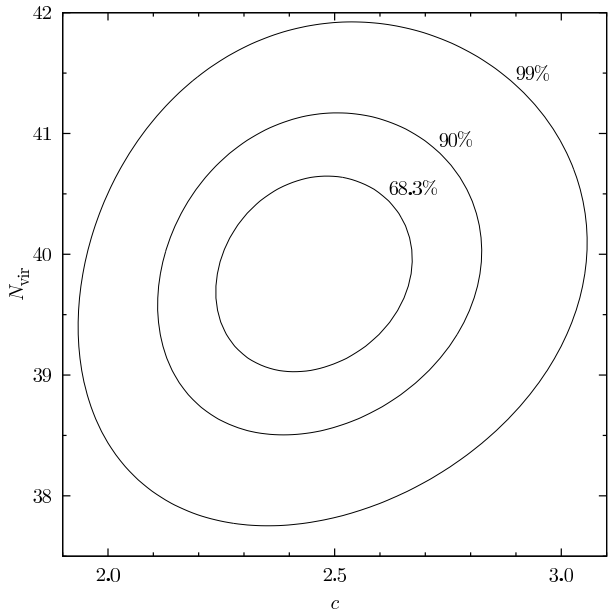}
  \caption{\label{fig.allgals.chisq} Contours of $\chi^2$ for the
  concentration parameter and amplitude of the NFW
  profile fitted to the projected galaxy density profile for all
  galaxies. Marginalized percentage confidence levels are
  indicated.}
\end{figure}

As discussed above, the reliability of the galaxy surface density
measurements deteriorates towards the centre of the groups. We
therefore restrict the profile fitting to points in the
range $0.05 < R/r_\rmn{vir} < 1.0$. We find projected NFW profiles
are capable of describing the observed profiles very well over the
radial range fitted, both for the separate blue and red galaxy
profiles and the combined galaxy profile. The best-fitting NFW profiles
are overlaid on the data in Figs \ref{fig.allgals.prof} and
\ref{fig.bytype.prof}, and the concentration parameters are given in
Table \ref{tab.nfw}. Fig. \ref{fig.allgals.chisq} shows the joint probability
distribution for the NFW concentration parameter and amplitude fitted
to the data for all galaxies; we find there to be no strong degeneracy between the two parameters.

\begin{table}
\caption{Best-fitting NFW concentration parameters. Errors are
  1-sigma, marginalized over the probability distribution of the other
  fitted parameter. $N_\rmn{vir}$ represents the average number of galaxies
  within $r_\rmn{vir}$ per group (above the minimum absolute magnitude $M_{b_\rmn{J},\rmn{com}} - 5 \log_\rmn{10}h = -17.5$).}
\label{tab.nfw}

\centering
\begin{tabular}{lccc}

\hline
& $N_\rmn{vir}$ & $c = r_\rmn{vir}/r_\rmn{s}$
\\
\hline
All galaxies & $39.8 \pm 0.8$ & $2.4 \pm 0.2$
\\
Red & $21.7 \pm 0.6$ & $3.9 \pm 0.5$
\\
Blue &$18.2 \pm 0.5$ & $1.3 \pm 0.2$
\\
\hline
\end{tabular}
\centering
\end{table}

It is a non-trivial assumption that stacking projected NFW profiles
with a range of concentration parameters should result in a composite
profile that is still NFW in shape, and which has a concentration
parameter indicative of the `average' of the underlying sample. We
therefore test this assumption in Appendix \ref{sec.stacking}. For a lognormal
distribution of $c$ we find that the composite profile retains the NFW
shape, but it has a concentration parameter which is lower than the
median $c$ of the underlying sample. The severity of the effect increases as the
distribution of $c$ becomes wider.

The best-fitting concentration parameters we give in Table \ref{tab.nfw} are broadly
consistent with the analyses of galaxy clusters in the $K$-band of the
2 Micron All-Sky Survey by \citet{lin} who obtain $c=2.90^{+0.21}_{-0.22}$, and of 14
CNOC clusters by \citet{carlberg} and \citet{marel} who find $c=3.7$
and $4.2$ respectively. \citet{hansen} find $c=$ 1--3 for
photometrically identified clusters in the SDSS. 

We caution that the close-pair incompleteness effects described above
mean that we cannot reliably constrain the radial distribution for $R< 0.05 r_\rmn{vir}$, so it is
impossible, for example, to determine whether the profile is cuspy or
flat-cored at its centre. Furthermore, a number of effects, some of
which we have already mentioned, are expected to
cause the measured concentration parameters to be depressed relative
to the true value:
\begin{enumerate}
\item stacking groups (Appendix \ref{sec.stacking}),
\item the uncertainty in the radial scaling of groups (Appendix \ref{sec.scaling}),
\item if we are still affected by close-pair incompleteness despite the angular
  diameter selection we have used, and
\item if there are discrepancies between the position of
  the `central' galaxy and the `true' group centre.
\end{enumerate}

\subsection{Mass dependence}
\label{sec.mass_dep}

The mean concentration parameter of dark matter halos in simulations
is found to be a decreasing function of mass, typically modelled as a
power law, $c = c_0 \left(M/M_*\right)^{-\lambda}$ with $\lambda \sim
0.1$ and $c_0 \sim 10$ (e.g. \citealt{bullock}).

\begin{figure}
  \includegraphics[width = 0.5\textwidth]{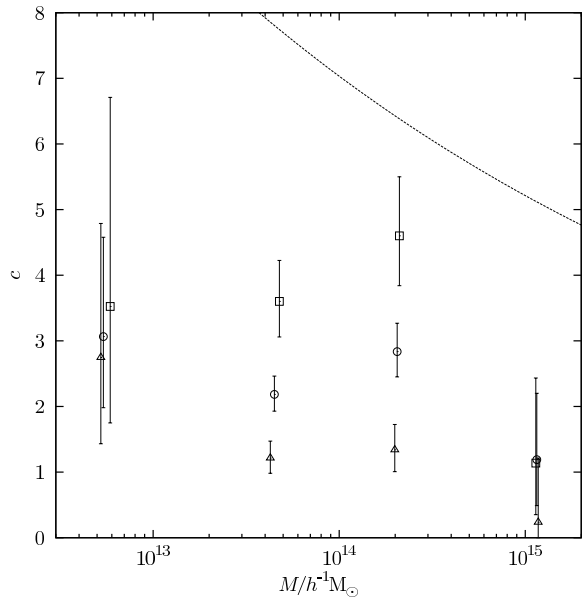}
  \caption{\label{fig.concs} Variation of group concentration
  parameter with mass, for red (squares), blue (circles), and all
  galaxies (triangles). For comparison, the dashed line shows the
  trend in the mean concentration parameter of dark matter haloes in
  simulations (due to \citealt{bullock}).}
\end{figure}

Fig. \ref{fig.concs} shows the measured concentration parameter
against group mass for red and blue galaxies separately, and
for the combined sample. These values were obtained by stacking groups within mass
bins, and fitting NFW profiles as above. The results suggest
a detection of a slight decrease in $c$ with mass, but the trend is
barely significant. When calculating halo model-based correlation
functions in the following section, we therefore choose to incorporate
no mass-dependence in the radial galaxy distribution; we confirm the
safety of this assumption below in Section \ref{sec.sensitive}.

The comparison with the results of \citet{bullock} reveals that, even
for the red galaxy distribution, we find the galaxy radial
distributions to be significantly less concentrated than those of
average dark matter haloes in simulations (but beware the caveats at
the end of Section \ref{sec.prof.fit}). The suggestion of a negative
correlation of concentration with group mass is in qualitative
agreement with the results of \citet{hansen}, who find $c$ to be a
decreasing function of galaxy occupation number.

% Correlation functions =============================================================

\section{Predicting correlation functions}
\label{sec.cfs}

Armed with the observations of the previous sections, we are now
equipped to calculate the halo model predictions for the
correlation function. \citet{hawkins} and \citet{madgwick_cf} have measured the two-point
correlation functions for the full 2dFGRS sample and for red and blue
subsamples. Care must be taken to ensure a fair, like-for-like
comparison, since the correlation function is known to depend on
redshift and luminosity (e.g. \citealt{zehavi}; \citealt{norberg_cf}). For this reason, the observed correlation functions are labelled by their
effective luminosity and effective redshift. These effective
quantities are pair-weighted measures, since the correlation function is
based on counting pairs of galaxies; hence, for example, the effective
redshifts -- $z_\rmn{s} = 0.15$ for the all-galaxy sample, and $z_\rmn{s} =
0.11$ for both the red and blue galaxy samples -- are somewhat higher
than the median redshifts of the samples. We use these
redshifts when calculating the halo model correlation functions for
the respective populations via the formalism of Section
\ref{sec.formalism}. 

In the case of the luminosity, the situation is
complicated by the fact that the 2dF samples are flux-limited, whereas the halo
model formalism effectively assumes volume-limiting. For a
volume-limited sample the effective luminosity is identical to the
mean luminosity. The minimum luminosity (required by
equation \ref{eqn.sel_fn}) is therefore determined by
solving
\begin{equation}
L_\rmn{s} = \frac{\int_{L_\rmn{min}}^{\infty}
  L~\phi(L)~\rmn{d}L}{\int_{L_\rmn{min}}^{\infty} \phi(L)~\rmn{d}L},
\end{equation}
for $L_\rmn{min}$. The effective luminosities of the samples used by
\citet{hawkins} and \citet{madgwick_cf} are, using  $M^{*} -
5\log_\rmn{10}h = -19.66$: all galaxies,
$L_\rmn{s}=1.4L_\rmn{*}$; red, $L_\rmn{s}=1.26L_\rmn{*}$; blue,
$L_\rmn{s}=0.95L_\rmn{*}$. The resulting limiting absolute magnitudes
are $M_{b_\rmn{J},\rmn{com}} - 5\log_\rmn{10}h = -19.25, -19.04,
-19.50$ for red, blue and all galaxies respectively. These limits have
been used in determining the halo occupation distribution of the 2PIGG
groups (Section \ref{sec.hod}), and the number densities of the galaxy
populations (Table \ref{tab.mcut}).

\begin{figure}
  \includegraphics[width = 0.5\textwidth]{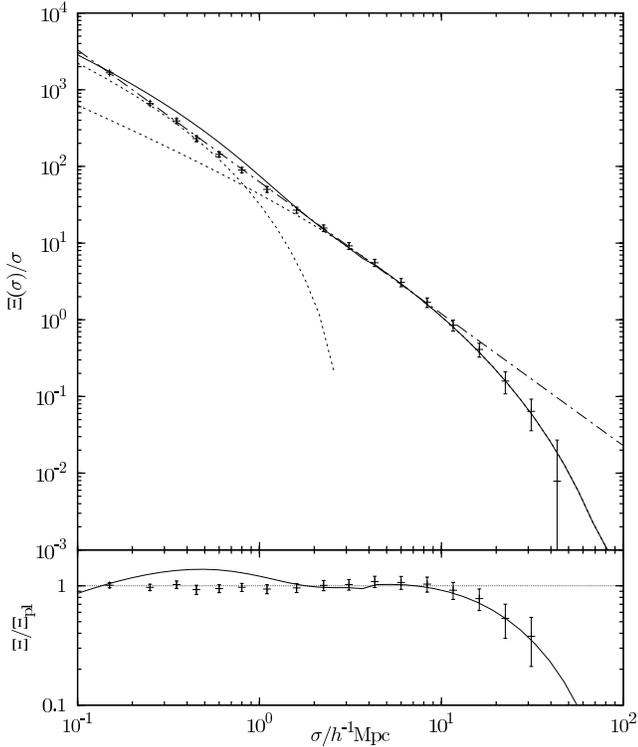}
  \caption{\label{fig.cf.allgals} Projected two-point correlation
  function for galaxies of all types. Data points represent the
  observed correlation function of \citet{hawkins} and the dot-dashed line is their
  power-law fit to these points. The solid line is our halo model-based
  prediction, and the short-dashed lines show the contributions from
  the individual one- and two-halo terms in the model. The lower panel
  shows the same data normalized by the power-law fit to the observations.}
\end{figure}

\begin{figure}
  \includegraphics[width = 0.5\textwidth]{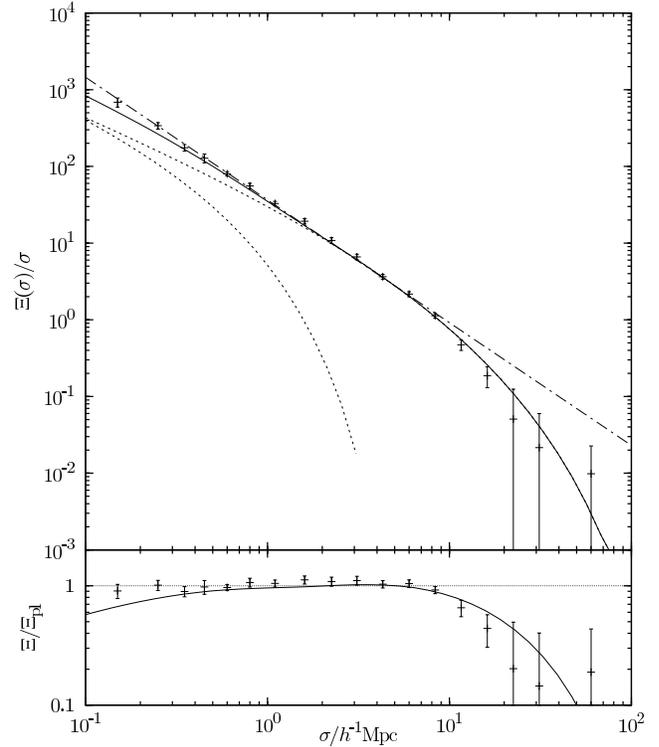}
  \caption{\label{fig.cf.late} As Fig. \ref{fig.cf.allgals}, but for
  blue galaxies only, and using the observed correlation function data of \citet{madgwick_cf}.}
\end{figure}

\begin{figure}
  \includegraphics[width = 0.5\textwidth]{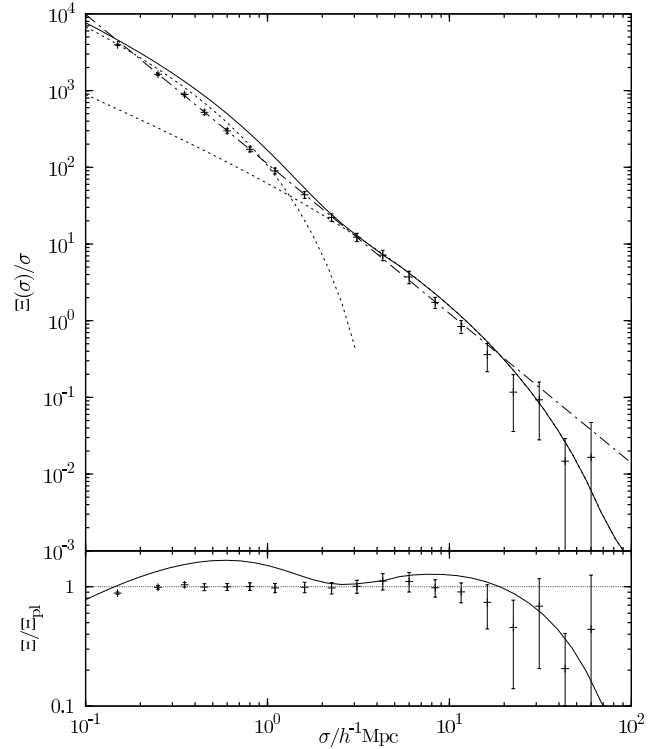}
  \caption{\label{fig.cf.early} As Fig. \ref{fig.cf.late}, but for red galaxies only.}
\end{figure}

In order to negate the effect of redshift distortions in the observed
correlation functions we use the radial projection,
\begin{equation}
\frac{\Xi(\sigma)}{\sigma} = \frac{2}{\sigma}
\int^{\infty}_{\sigma}\frac{\xi(r)~r~\rmn{d}r}{(r^2 - \sigma^2)^{1/2}}~.
\end{equation}
In Figs \ref{fig.cf.allgals}, \ref{fig.cf.late}
and \ref{fig.cf.early} we compare the halo model predictions with the observed projected
correlation functions.

In producing these halo model correlation functions we have simply used the
best-fitting halo profiles and occupation
distributions, deliberately making no attempt to fit to the observed
correlation function
data. We find that the model successfully recovers the approximate shape and amplitude
of the observed correlation functions across the range of distance
scales considered. In particular, the model correlation functions
trace the obvious divergence from the power-law form at large scales. The
different observed HODs and halo profiles of red and
blue galaxies derived in Sections \ref{sec.hod} and \ref{sec.rad.dist}
are seen to give rise to the expected relative strengths of
correlations on small scales, although the halo model predictions do show significant
deviations from the power-law form on these scales.

\subsection{Galaxy bias on large scales}
\label{sec.bias}

On large scales the differences in the correlation functions of the
different galaxy classes are
encapsulated in the galaxy bias parameter (equation
\ref{eqn.bias_param}). To allow comparison with other results, we
correct the galaxy bias parameter for redshift and luminosity
dependence via
\begin{equation}
b_\rmn{gal}(L, 0) = b_\rmn{gal}(L, z_\rmn{s}) D(z_\rmn{s})
\end{equation}
and
\begin{equation}
b_\rmn{gal}(L,0)/b_\rmn{gal}(L_\rmn{*},0) = 0.85 + 0.15(L/L_\rmn{*}).
\end{equation}
The first relation follows from the assumption that galaxy clustering
evolves weakly over this redshift range (i.e. $\sigma_\rmn{8, gal}$ remains
approximately constant). The second is found from
correlation function analysis by \citet{norberg_bias}. The predicted galaxy bias
factors are in Table \ref{tab.bias}. \citet{lahav} have measured the galaxy
bias factor by combining the 2dFGRS galaxy power spectrum with the
post-WMAP CMB data and found
\begin{equation}
b_\rmn{gal}(L_\rmn{*}, 0) \approx 0.96 \pm 0.08,
\end{equation}
in good agreement with
our predicted value for all galaxies: $b_\rmn{gal}(L_\rmn{*}, 0) = 0.92$.

Furthermore, the relative bias between our red and blue samples,
$b_\rmn{red}/b_\rmn{blue} = 1.4$, is in excellent
agreement with that found by \citet{cole}, who measure the power
spectra for red and blue 2dFGRS subsamples defined by a colour cut at $\rmn{b}_\rmn{J} -
r_\rmn{F} = 1.07$. 

\begin{table}
\caption{Galaxy bias parameters predicted by the halo model. Note that
the effective luminosities and redshifts differ between the three
galaxy samples.}
\label{tab.bias}

\centering
\begin{tabular}{lccc}

\hline
& $b_\rmn{gal}(L_\rmn{s}, z_\rmn{s})$ & $b_\rmn{gal}(L_\rmn{*}, 0)$
\\
\hline
All galaxies & $1.05$ & $0.92$
\\
Red & $1.23$ & $1.12$
\\
Blue & $0.85$ & $0.81$
\\
\hline
\end{tabular}
\centering
\end{table}

\subsection{Sensitivity to parameters}
\label{sec.sensitive}

The \emph{halo} bias factor, $b(M)$, (equation
\ref{eqn.halo_bias}) is a monotonically increasing function of mass,
with the lowest mass haloes in fact being \emph{anti-biased}. The
galaxy bias factor predicted by the halo model (equation
\ref{eqn.gal_bias}) will therefore be controlled by the relative occupation of
high to low mass groups: thus, since the red galaxies have a steeper
$\langle N|M \rangle$ than the blue, they also have a greater
large-scale bias factor.

The differences on small scales are, in principle, due to the different
radial profiles as well as the relative halo occupation numbers of the
galaxy classes. However, it turns out that on the scales we are
currently able to access observationally (and given the observational results
of the previous sections) the correlation function is much
more sensitive to the differences in the halo occupation
distribution (see also \citealt{sheth_cf}; \citealt{berlind_weinberg}; \citealt{scranton}). For
example, substituting the dark matter concentration parameter relation
of \citet[see Section \ref{sec.mass_dep}]{bullock}, in place of the
observed values has a perceptible effect only on scales $r \la 3
\times 10^{-1} \rmn{Mpc}/h$.

\begin{figure}
  \includegraphics[width = 0.5\textwidth]{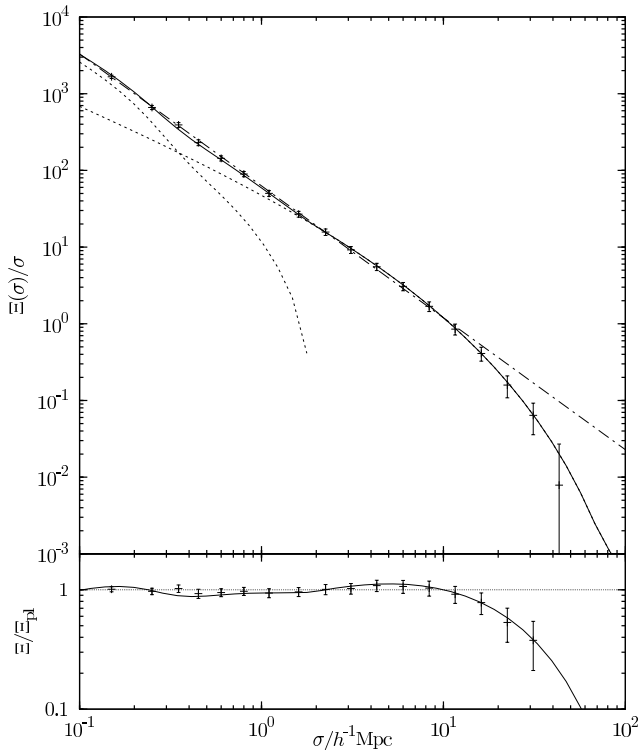}
  \caption{\label{fig.cf.tweaked} As Fig. \ref{fig.cf.allgals}, but
  the halo model predictions are based on the more sophisticated HOD
  described by equations \ref{eqn.soph_hod} and
  \ref{eqn.sub_pois}. This form of the HOD better reproduces the
  power-law on small scales, while still being consistent with the
  observational constraints of Section \ref{sec.hod}.}
\end{figure}

As we explain in Section \ref{sec.lowmass}, the HOD for low mass
groups is unconstrained by the observations. In consequence, we are
forced to assume an extrapolation of the high mass HOD to low
masses. The shape of the intra-halo term of the correlation function
is strongly dependent on the particular form assumed for the halo
occupation distribution at low group masses, but the overall amplitude is robust to
such details, being dominated by the contribution from galaxy pairs in
high-mass haloes. \citealt{berlind_weinberg} demonstrate the
contributions to $\xi(r)$ from galaxy pairs as a function of halo
mass: in general lower mass haloes become influential on smaller
scales. By using a more sophisticated extrapolation to low masses it
is possible to produce a halo model correlation function which more
faithfully reproduces the power-law form on small scales. For example,
we show in Fig. \ref{fig.cf.tweaked} the correlation function for all
galaxies produced
by using a halo occupation distribution of the form
\begin{equation}
  \label{eqn.soph_hod}
  \langle N | M \rangle = \left\{ \begin{array}{ll}
    \left(M/M_0\right)^\beta & (M \ge M_\rmn{0}) \\
    1 & (M_\rmn{cut} \le M < M_\rmn{0}) \\
    0 & (M < M_\rmn{cut}) 
  \end{array} \right.
\end{equation}
with $M_\rmn{0}$ and $\beta$ as given in Table \ref{tab.hod}, 
and introducing an ad hoc sub-Poisson distribution at low masses:
\begin{equation}
  \label{eqn.sub_pois}
  \alpha(M) = \left\{ \begin{array}{ll}
    1 & (M \ge M_\rmn{0}) \\
    0.2 & (M < M_\rmn{0})
  \end{array} \right.~.
\end{equation}
As described in Section \ref{sec.lowmass}, we constrain $M_\rmn{cut}$ by matching the observed number
density. For this choice of HOD we obtain a much
more realistic $M_\rmn{cut}
= 1.4 \times 10^{12} h^{-1}\rmn{M}_{\sun}$. Both this and our simplest
possible HOD (equation \ref{eqn.hod}) are entirely consistent
with our observations, and are successful at recovering the amplitude
of the correlation function. However, the more sophisticated form
gives a much better fit to the observations on small scales.

\section{Discussion}

Motivated by the halo model, we have used the 2dF Percolation-Inferred
Galaxy Group catalague to make direct observational estimates of the
distribution functions of galaxies within and amongst haloes. We have
found a projected NFW profile provides a good fit to the composite
projected surface density of galaxies in red and blue subclasses, and
for the combined sample. The radial distribution of the red galaxies
was found to be the most concentrated towards the group centre, but,
even for these, the concentration is significantly lower than for
dark matter haloes measured in simulations.

We have shown that at high masses the mean occupation number of the
2PIGGs with red, blue, and all galaxies can be described by a
power-law in the mass, with the number of red galaxies increasing
most steeply with the mass. The difficulty of obtaining reliable
dynamical mass estimates for groups with low occupation numbers made
it difficult to constrain the halo occupation distribution in this
regime.

These observations have allowed us to test the halo model against fully
self-consistent observations: the same 2dFGRS galaxy populations used
to compile the 2PIGG catalogue have previously had their correlation
functions directly determined. Despite the uncertainty regarding the
form of the halo occupation distribution at low halo masses, the halo
model proves successful at reproducing the near power-law form of
the galaxy correlation functions. Most impressively, the model also
succesfully recovers the relative biasing of the red and blue subpopulations.
In agreement with \citet{sheth_cf}, we identify the halo occupation
number, $\langle N | M \rangle$, as the
dominant factor contributing to the differences in the correlation
functions of red and blue galaxies: the HOD for red galaxies gives a
greater weighting to high-mass haloes leading to a
stronger clustering signal on all scales. The differences in the radial distributions
of the galaxies in groups are found to be relatively unimportant on
the distance scales we are presently able to access observationally,
but do begin to exert influence at the smallest scales considered.

There are of course any number of forms which could be chosen for the extrapolation of the
HOD to low mass (see for example, \citealt{kravtsov};
\citealt{diaferio}; \citealt{scranton}; \citealt{maglio}). We have
shown that the power-law form of the observed correlation functions
does place constraints on this extrapolation. However, in the
intra-halo term there is certainly degeneracy between $\alpha(M)$ and
$\langle N | M \rangle$, so it is unlikely that the two-point
correlation can be used to constrain a unique solution. Direct
measurement of the HOD for low-mass haloes requires much improved mass
estimates, and represents a difficult observational challenge. Higher
order clustering statistics represent a more stringent test for the
halo model, and offer the possibility of indirectly
constraining the low-mass HOD in a more model-independent fashion than
is possible from the two-point function alone (e.g. \citealt{ma};
\citealt{scoccimarro}; \citealt{wang}; \citealt*{fosalba}).

\section*{Acknowledgments}

We gratefully acknowledge the assistance of Vince Eke, Josh
Frieman, Yehuda Hoffman and Frank van den Bosch. We also thank our
referee, Ravi Sheth, for many useful suggestions. AAC is supported
by an Isle of Man Department of Education
Postgraduate Studies Grant. OL acknowledges a PPARC Senior Research
Fellowship.

% Bibliography ===============================================================

\appendix

\section{Stacking groups}
\label{sec.stacking}

In order to allow accurate measurement of the galaxy radial
distribution it is necessary to combine many groups so as to improve
the statistics. Even the largest groups in the 2PIGG catalogue do not have sufficient galaxy
members to allow the concentration parameter to be well constrained
for individual groups.

The groups are scaled by their virial radius. However, NFW profiles
with different concentration parameters are not self-similar when
scaled in this way (scaling by $r_\rmn{s}$, on the other hand, does produce
self-similar profiles, but this approach is not practical since
estimating $r_\rmn{s}$ is non-trivial).
It is not clear therefore that the resulting profile will retain the NFW form, or
that the concentration parameter thus obtained will be at all
indicative of the `average' concentration of the groups in the sample.

We do not have any indication of the expected distribution of the
concentration parameter for galaxy groups, due to the difficulty in
measuring the profile for individual groups. For dark matter haloes,
simulations (e.g. \citealt{jing}; \citealt{bullock}) suggest that
$c$ has a lognormal distribution at fixed halo mass,
\begin{equation}
p(c|M)\rmn{d}c = \frac{1}{\sqrt{2\pi}\sigma_c}\exp{\left[-\frac{(\ln{c}
      - \ln\bar{c})^2}{2\sigma^2_c}\right]}\rmn{d}\ln{c}
\end{equation}
where the median, $\bar{c}$, contains the only dependence on halo
mass (see \S\ref{sec.mass}). Both \citet{jing} and \citet{bullock} find $\sigma_c \sim 0.2$.
Although there is no reason to assume that these results apply equally well
to the galaxy group profiles, we use this distribution in order to
illustrate the effect of stacking profiles with a wide range of
concentrations.

Stacking profiles drawn from a distribution $p(c)$ will produce the
aggregate profile,
\begin{equation}
\bar{\Sigma}(R) = \int \Sigma(R|c)~p(c)~\rmn{d}c.
\end{equation}
We have computed this aggregate profile for a range of lognormal
distributions, and fitted a projected NFW profile to the result, just
as for the real data (i.e. using only the region $0.05 <
R/r_\rmn{vir} < 1.0$).

The stacked profiles can be well fitted by the NFW profile, even for
the widest distributions. However, the concentration parameter of the
best-fitting profile is systematically lower than the median of the
underlying distribution. Table \ref{tab.stacking} shows the results
for the range of distributions considered.

The disparity between the median and fitted $c$ becomes increasingly significant as
the distribution broadens. However, the best-fitting $c$ is found to be a fair
indicator for the \emph{mode} (i.e. peak position) of the underlying
lognormal distribution, particularly when the dispersion is relatively low.

\begin{figure}
  \includegraphics[width = 0.5\textwidth]{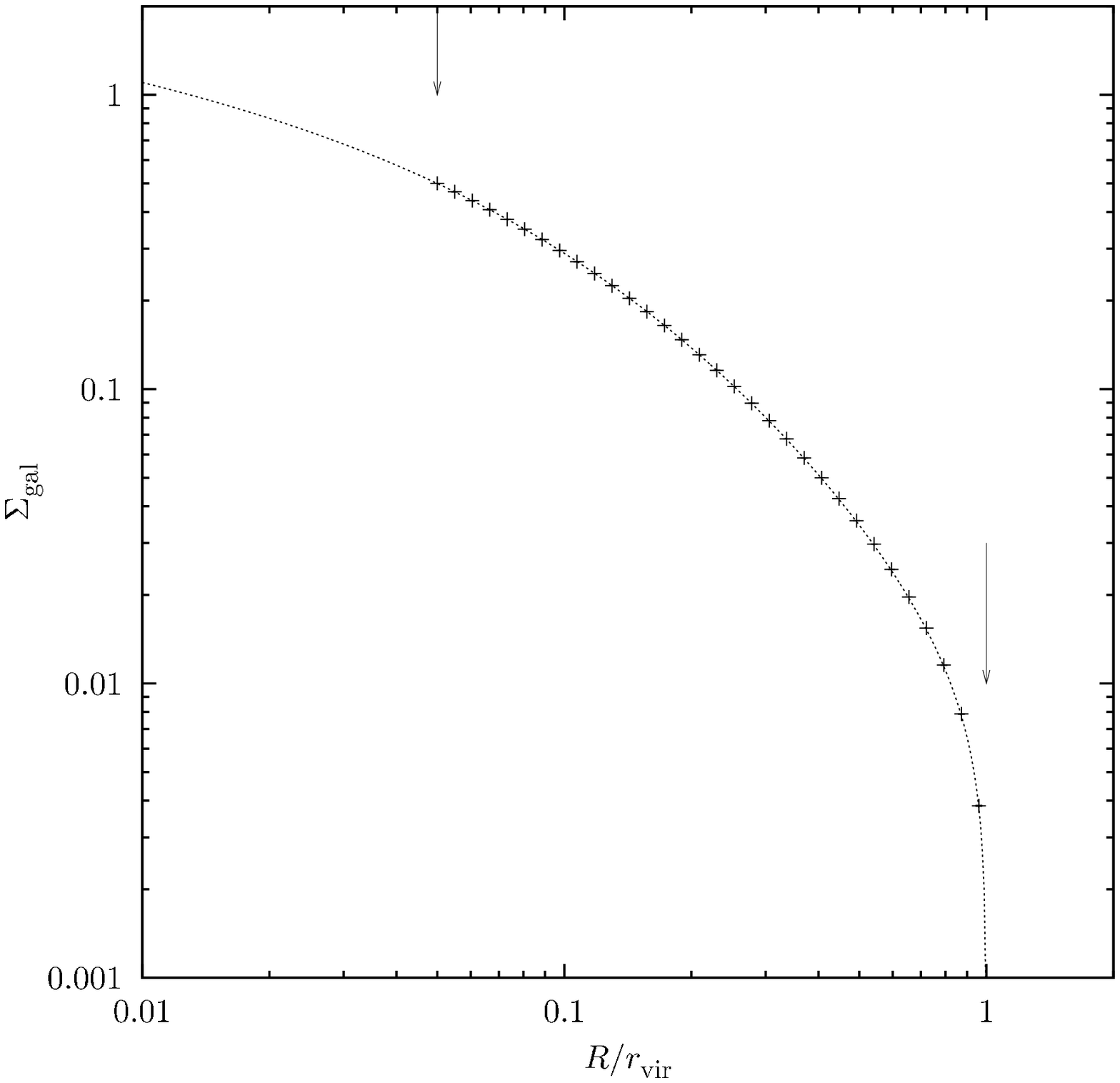}
  \caption{\label{fig.stack_fit} The crosses show the simulted aggregate
  profile obtained by stacking NFW profiles drawn from a lognormal
  distribution with median $\bar{c} = 5$ and width $\sigma_c =
  0.2$. The dotted line is the NFW profile which best fits the stacked
  points; it has $c = 4.66$.}
\end{figure}

\begin{table}
\caption{Results from fitting NFW profiles to stacked profiles having
  concentration parameters drawn from lognormal
  distributions with the given median ($\bar{c}$) and width
  ($\sigma_c$). The concentration parameter of the best-fitting NFW
  profile, $c_\rmn{fit}$, is found
  to be a fair indicator for the \emph{mode} of the underlying distribution.}
\label{tab.stacking}

\centering
\begin{tabular}{lccc}

\hline
$\bar{c}$ & $\sigma_c$ & $c_\rmn{fit}$ & $c_\rmn{mode}$
\\
\hline
10  &  0.1  &  9.79  &  9.90 \\
10  &  0.2  &  9.19  &  9.61 \\
10  &  0.5  &  6.18  &  7.79 \\
10  &  1.0  &  2.22  &  3.68 \\
5  &  0.1  &  4.91  &  4.95 \\
5  &  0.2  &  4.66  &  4.80 \\
5  &  0.5  &  3.31  &  3.90 \\
5  &  1.0  &  1.34  &  1.84 \\

\hline
\end{tabular}
\centering
\end{table}

\subsection{Mass dependence}
\label{sec.mass}

\citet{bullock} find the median concentration parameter for dark
matter haloes to be a
function of halo mass: $\bar{c}(M) \approx 9 (M/M_*)^{-0.13}$ at $z=0$. In
combination with the mass function, $n(M)$ (equation \ref{eqn.mf}), this allows us to
write down the global distribution of $c$ as
\begin{equation}
\label{eqn.globalc}
f(c) = \int p(c|M)~n(M)~\rmn{d}M .
\end{equation}

\begin{figure}
  \includegraphics[width = 0.5\textwidth]{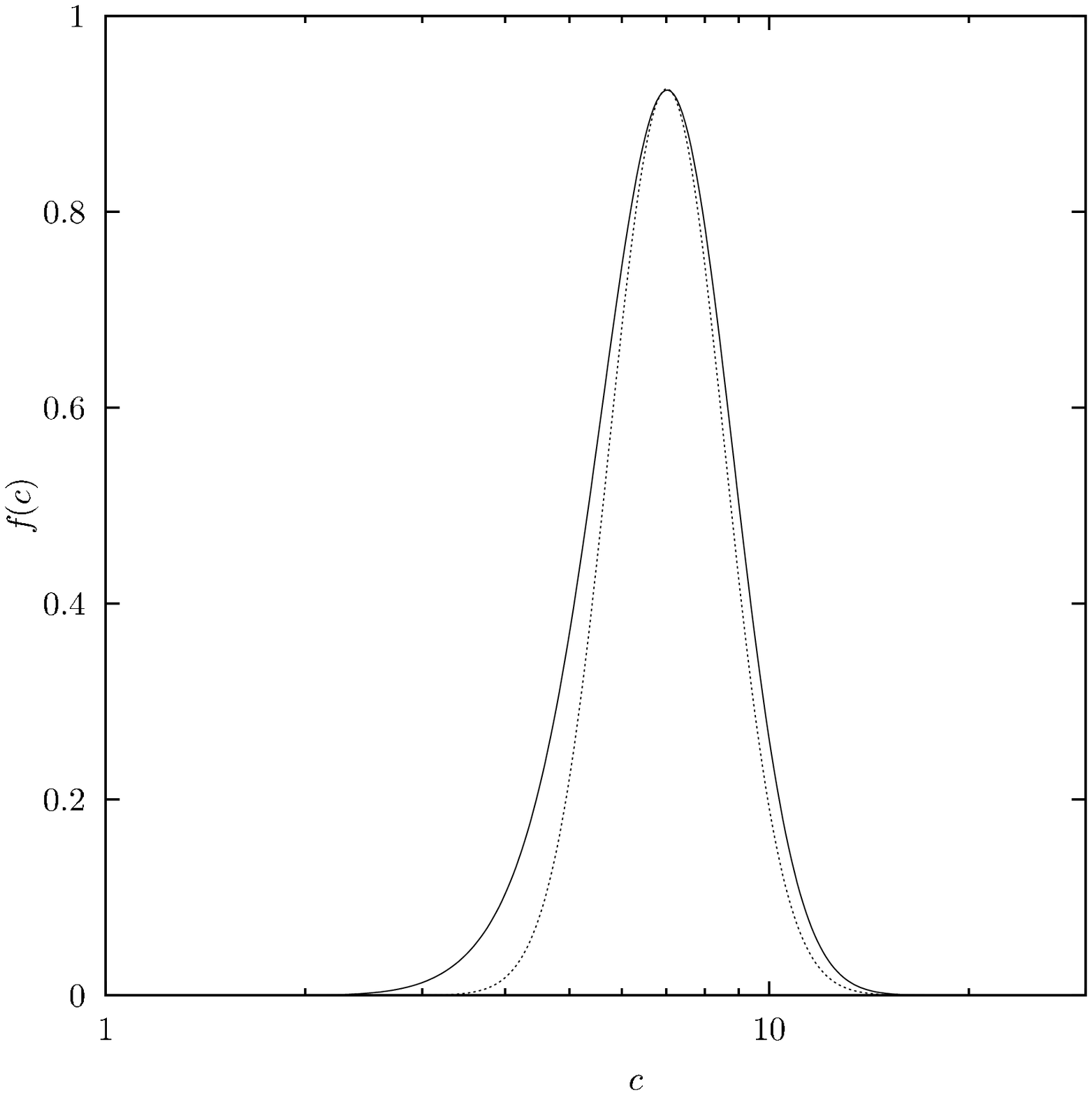}
  \caption{\label{fig.global} The solid curve is the global distribution of $c$ for dark
  matter haloes (equation \ref{eqn.globalc}). For comparison, the dotted line shows a lognormal distribution with
  width $\sigma_c = 0.2$ and having the same mode as the global
  distribution.}
\end{figure}

Figure \ref{fig.global} shows this global distribution for haloes in
the mass range $10^{13}<M/h^{-1}\rmn{M}_{\sun} < 10^{16}$, assuming
$p(c|M)$ is lognormal with $\sigma_c = 0.2$. The combination with the
mass function leads to a slight widening of the distribution, particularly
on the low-$c$ tail.

This result clearly depends on the assumption of forms for
$\bar{c}(M)$ and $p(c|M)$, both of which we have derived from
simulated dark matter haloes. Our observational results (Fig. \ref{fig.concs}) suggest that
$\bar{c}_\rmn{gal}$ depends on the host halo mass no more strongly than does
the dark matter concentration parameter.

What is most important is that the combination with the mass function has not
significantly altered the shape of the distribution, so that the
results of Section \ref{sec.stacking} still hold when we stack groups
covering a wide range of masses, albeit with a slightly wider underlying
distribution of $c$.

\subsection{Scaling uncertainty}
\label{sec.scaling}

The dynamical masses determined by \cite{ekea} for the 2PIGG groups
have a large scatter, particularly for small (low membership) groups. Figure 3 of
\citet{ekeb} illustrates the dependence of this scatter on group
membership. Since low-membership groups are the most abundant in the
2PIGG catalogue, one might expect the scaling error to be dominated by
their contribution. Fortunately however, the impact on the stacked profile of groups with a
particular membership, $N$, is proportional to $N \times n(N)$. Thus,
although there are many more low-membership groups, the individual
groups have proportionally less
impact on the stacked profile than a single high-membership group. It turns
out that $N \times n(N)$ is nearly independent of $N$ for the 2PIGG
catalogue, so the overall scaling error receives roughly equal
contributions from all group memberships. 

\begin{figure}
  \includegraphics[width = 0.5\textwidth]{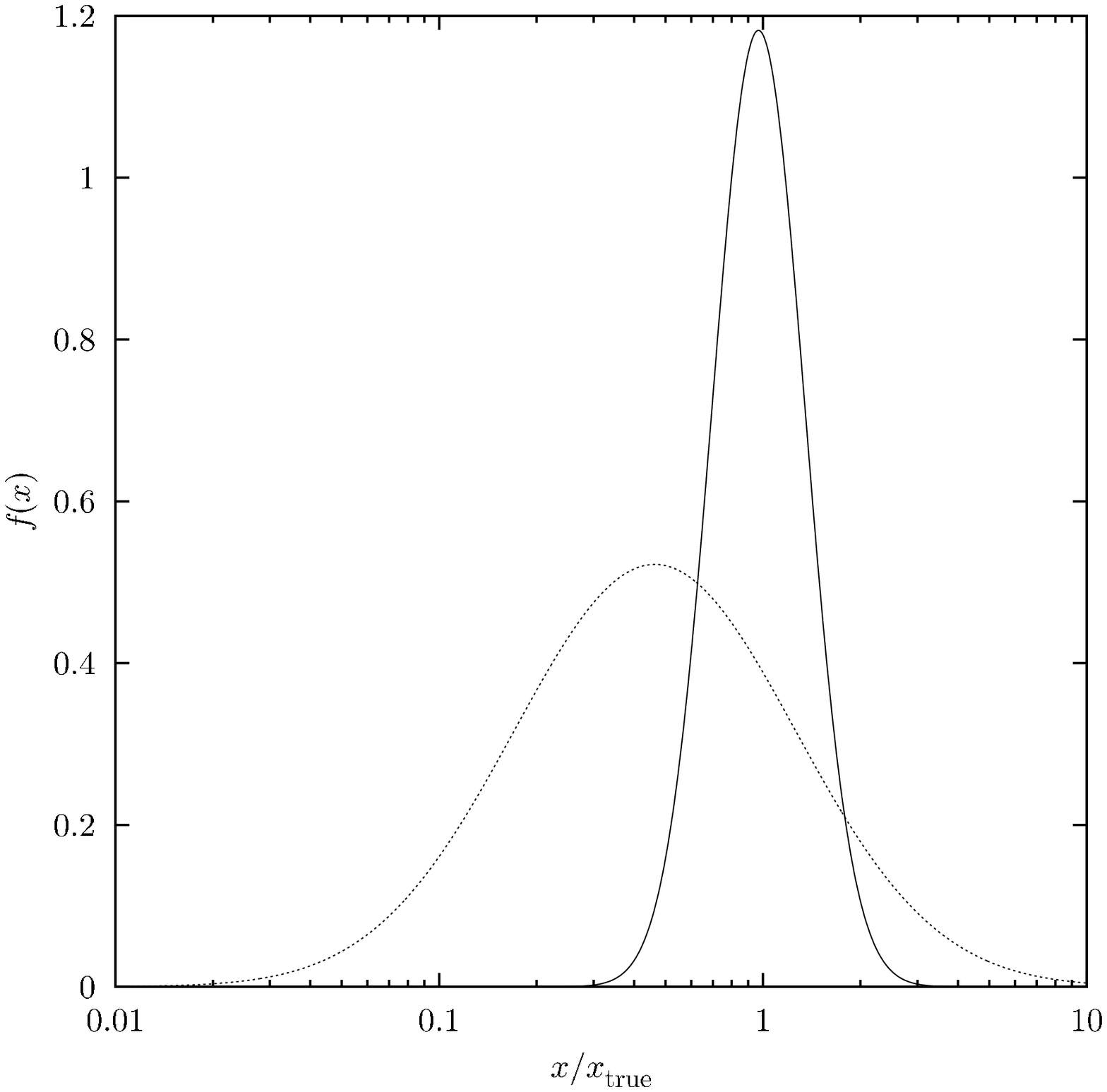}
  \caption{\label{fig.blur} The dotted line is the error distribution
  in the group mass estimates (derived from equation
  \ref{eqn.2pigg_mass}) for groups having $N=10$ observed galaxy
  members. Through equation \ref{eqn.rvir} this implies an error
  distribution in the virial radius, shown by the solid line.}
\end{figure}

We adopt $N=10$ as our fiducial point, since the error distribution
here is close to average. The distribution of $M/M_\rmn{true}$ at
$N=10$ is roughly lognormal with median 1.26 and $\sigma \sim
1.0$. Propagating the error distribution of $M/M_\rmn{true}$ through
equation \ref{eqn.rvir} implies that $r/r_\rmn{true}$ has a lognormal distribution
with median 1.08 and $\sigma \sim 0.33$: the $r_\rmn{vir} \propto
M^{1/3}$ relation means the large scatter in the mass estimates is
considerably suppressed, leading to a somewhat narrower distribution
of $r/r_\rmn{true}$ (Fig. \ref{fig.blur}).

\begin{figure}
  \includegraphics[width = 0.5\textwidth]{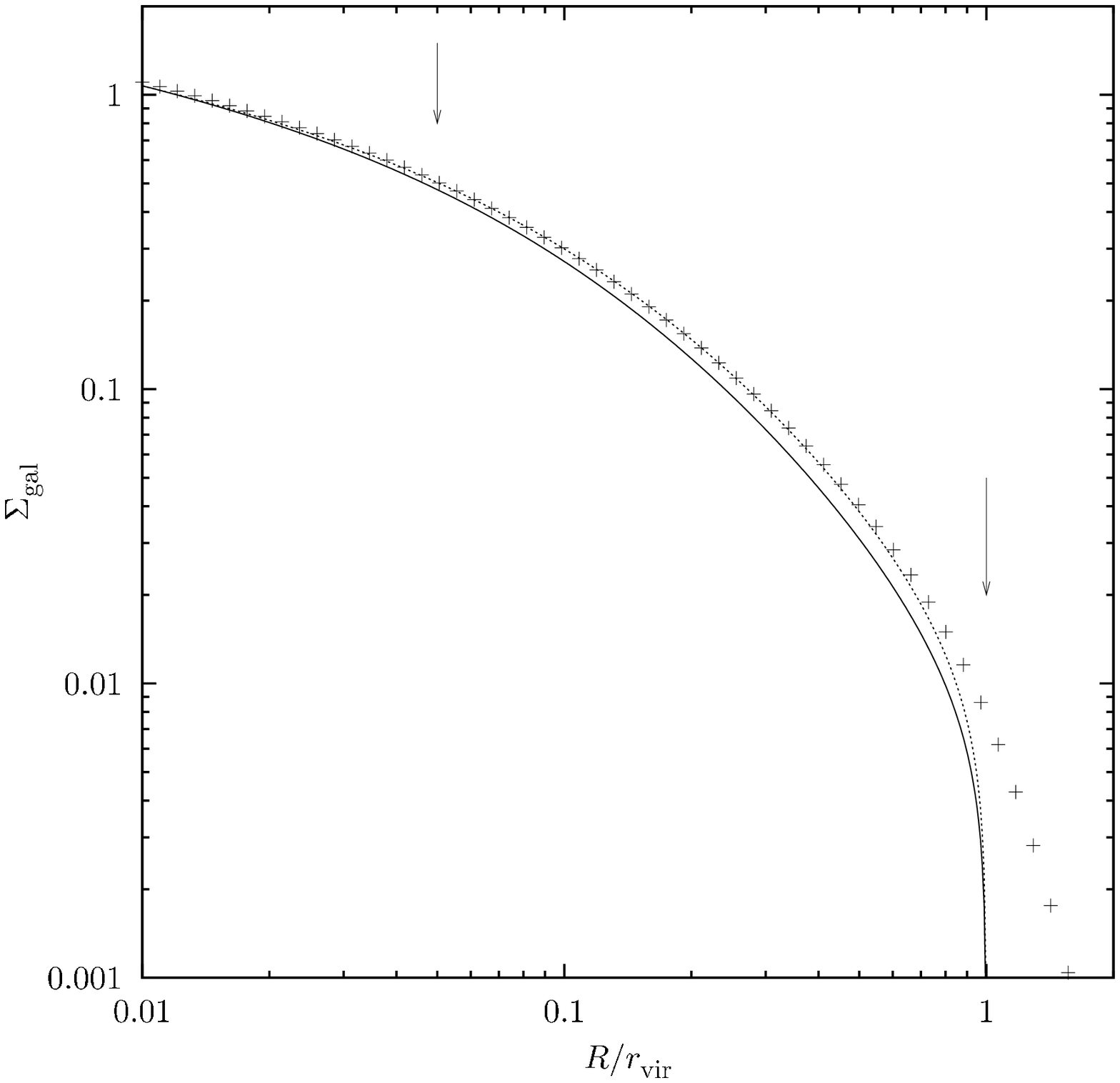}
  \caption{\label{fig.blur_prof} Effect of scaling uncertainty on the
  aggregate profile obtained by stacking groups. The solid line shows
  a projected NFW profile with $c=5.0$. Crosses show simulated data
  generated by equation \ref{eqn.scaling},
  assuming the scaling uncertainty described in the text. The dotted
  line is the NFW profile which best-fits the data; it has $c = 4.13$.}
\end{figure}

The aggregate profile obtained by stacking profiles with an uncertain
radial scaling is given by
\begin{equation}
\label{eqn.scaling}
\tilde{\Sigma}(R) = \int \Sigma\left(\frac{R}{s}~|~c\right) p(s) \rmn{d}s,
\end{equation}
where we have defined the scaling error, $s = r/r_\rmn{true}$, and its
probability distribution $p(s)$ which, as we have discussed above, may
be approximated by a lognormal distribution. This profile is illustrated in
Fig. \ref{fig.blur_prof}.

The scaling uncertainty has the effect of smoothing the cut-off at the
virial radius, just as observed in Section \ref{sec.rad.dist}, and causes a
significant reduction in the measured concentration parameter.

\label{lastpage}


\begin{thebibliography}{}
  \bibitem[\protect\citeauthoryear{Bartelmann}{1996}]{bartelmann}
    Bartelmann M., 1996, A\&A, 313, 697

  \bibitem[\protect\citeauthoryear{Benson et~al.}{2000}]{benson}
    Benson A.~J., Cole S., Frenk C.~S., Baugh C.~M., Lacey C.~G.,
    2001, MNRAS, 327, 1041

  \bibitem[\protect\citeauthoryear{Berlind \&
  Weinberg}{2002}]{berlind_weinberg}
    Berlind A.~A., Weinberg D.~H., 2002, ApJ, 575, 587

  \bibitem[\protect\citeauthoryear{Bullock et~al.}{2001}]{bullock}
    Bullock, J.~S., Kolatt, T.~S., Sigad, Y., Somerville, R.~S.,
    Kravtsov, A.~V., Klypin, A.~A., Primack, J.~R., Dekel, A., 2001,
    MNRAS, 321, 559
  
  \bibitem[\protect\citeauthoryear{Carlberg et~al.}{1997}]{carlberg}
    Carlberg R.~G., et~al., 1997, ApJ, 485, L13

  \bibitem[\protect\citeauthoryear{Casas-Miranda
      et~al.}{2002}]{casas-miranda}
    Casas-Miranda R., Mo H.~J., Sheth R.~K., Boerner G., 2002, MNRAS,
    333, 730

  \bibitem[\protect\citeauthoryear{Cole et~al.}{2005}]{cole}
    Cole S. \& The 2dFGRS team, 2005, preprint (astro-ph/0501174)

  \bibitem[\protect\citeauthoryear{Cooray \& Sheth}{2002}]{cooray}
    Cooray A., Sheth R.~K., 2002, Phys. Rep., 371, 1

  \bibitem[\protect\citeauthoryear{Croton et~al.}{2005}]{croton}
    Croton D.~J. \& The 2dFGRS team, 2005, MNRAS, 356, 1155

  \bibitem[\protect\citeauthoryear{Dressler}{1980}]{dressler}
    Dressler A., 1980, ApJ, 236, 351

  \bibitem[\protect\citeauthoryear{Efstathiou
  et~al.}{2002}]{efstathiou}
    Efstathiou G. \& The 2dFGRS team, 2002, MNRAS, 330, 29

  \bibitem[\protect\citeauthoryear{Eke, Cole \& Frenk}{Eke et
  al.}{1996}]{eke_delta}
    Eke V.~R., Cole S., Frenk C.~S., 1996, MNRAS, 282, 263

  \bibitem[\protect\citeauthoryear{Eke et~al.}{2004a}]{ekea}
    Eke V.~R. \& The 2dFGRS team, 2004a, MNRAS, 348, 866

  \bibitem[\protect\citeauthoryear{Eke et~al.}{2004b}]{ekeb}
    Eke V.~R. \& The 2dFGRS team, 2004b, MNRAS, 355, 769
    
  \bibitem[\protect\citeauthoryear{Fosalba, Pan \& Szapudi}{Fosalba
  et~al.}{2005}]{fosalba}
    Fosalba P., Pan J., Szapudi I., 2005, preprint (astro-ph/0504305)

  \bibitem[\protect\citeauthoryear{Goto et~al.}{2003}]{goto}
    Goto T., Yamauchi C., Fujita Y., Okamura S., Sekiguchi M., Smail I., Bernardi M., Gomez P. L., 2003, MNRAS, 346, 601

  \bibitem[\protect\citeauthoryear{Hansen et~al.}{2004}]{hansen}
    Hansen S.~M., McKay T.~A., Wechsler R.~H., Annis J., Sheldon
    E.~S., Kimball A., 2004, preprint (astro-ph/0410467)

  \bibitem[\protect\citeauthoryear{Hawkins et~al.}{2003}]{hawkins}
    Hawkins E. \& The 2dFGRS team, 2003, MNRAS, 346, 78
  
  \bibitem[\protect\citeauthoryear{Jing}{2000}]{jing}
    Jing Y.~P., 2000, ApJ, 535, 30

  \bibitem[\protect\citeauthoryear{Kauffmann et~al.}{1999}]{kauffmann}
    Kauffmann G., Colberg J.~M., Diaferio A., White S.~D.~M., 1999,
    MNRAS, 303, 188

  \bibitem[\protect\citeauthoryear{Kravtsov et~al.}{2004}]{kravtsov}
    Kravtsov A.~V., Berlind A.~A., Wechsler R.~H., Klypin A.~A.,
    Gottl\"ober S., Allgood B., Primack J.~R., 2004, ApJ, 609, 35

  \bibitem[\protect\citeauthoryear{Lahav et~al.}{2002}]{lahav}
    Lahav O. \& The 2dFGRS team, 2002, MNRAS, 333, 961

  \bibitem[\protect\citeauthoryear{Lin, Mohr \& Stanford}{Lin
  et~al.}{2004}]{lin}
    Lin Y.-T., Mohr J.~J., Stanford S.~A., 2004, ApJ, 610, 745

  \bibitem[\protect\citeauthoryear{Ma \& Fry}{2000}]{ma}
    Ma C., Fry J.~N., 2000, ApJ, 543, 503

  \bibitem[\protect\citeauthoryear{Madgwick et~al.}{2002}]{madgwick_eta}
    Madgwick D. S. \& The 2dFGRS team, 2002, MNRAS, 333, 133

  \bibitem[\protect\citeauthoryear{Madgwick et~al.}{2003}]{madgwick_cf}
    Madgwick D. S. \& The 2dFGRS team, 2003, MNRAS, 344, 847

  \bibitem[\protect\citeauthoryear{Magliocchetti \&
  Porciani}{2003}]{maglio}
    Magliocchetti M., Porciani C., 2003, MNRAS, 346, 186

  \bibitem[\protect\citeauthoryear{Mo \& White}{1996}]{mo_white}
    Mo H.~J., White S.~D.~M., 1996, MNRAS, 282, 347

  \bibitem[\protect\citeauthoryear{Nagai \& Kravtsov}{2005}]{nagai}
    Nagai D., Kravtsov A.~V., 2005, ApJ, 618, 557

  \bibitem[\protect\citeauthoryear{Navarro, Frenk \& White}{Navarro et~al.}{1997}]{nfw}
    Navarro J.~F., Frenk C.~S., White S.~D.~M., 1997, ApJ, 490, 493

  \bibitem[\protect\citeauthoryear{Neyman \&
  Scott}{1952}]{neyman_scott}
    Neyman J., Scott E.~L., 1952, ApJ, 116,144

  \bibitem[\protect\citeauthoryear{Norberg
  et~al.}{2001}]{norberg_bias}
    Norberg P. \& The 2dFGRS team, 2001, MNRAS, 328, 64

  \bibitem[\protect\citeauthoryear{Norberg et~al.}{2002a}]{norberg_cf}
    Norberg P. \& The 2dFGRS team, 2002a, MNRAS, 332, 827

  \bibitem[\protect\citeauthoryear{Norberg et~al.}{2002b}]{norberg_lf}
    Norberg P. \& The 2dFGRS team, 2002b, MNRAS, 336, 907

  \bibitem[\protect\citeauthoryear{Padilla et~al.}{2004}]{padilla}
    Padilla N.~D. \& The 2dFGRS team, 2004, MNRAS, 352, 211

  \bibitem[\protect\citeauthoryear{Peacock \& Smith}{2000}]{peacock}
    Peacock J.~A., Smith R.~E., 2000, MNRAS, 318, 1144

  \bibitem[\protect\citeauthoryear{Peebles}{1980}]{peebles}
    Peebles P.~J.~E., 1980, The Large Scale Structure of the
    Universe. Princeton Univ. Press, Princeton NJ

  \bibitem[\protect\citeauthoryear{Percival et~al.}{2001}]{percival}
    Percival W.~J. \& The 2dFGRS team, 2001, MNRAS, 327, 1297

  \bibitem[\protect\citeauthoryear{Postman \& Geller}{1984}]{postman}
    Postman M., Geller M.~J., 1984, ApJ, 281, 95

  \bibitem[\protect\citeauthoryear{Press \&
  Schechter}{1974}]{press_schechter}
    Press W.~H., Schechter P., 1974, ApJ, 187, 425

  \bibitem[\protect\citeauthoryear{Scherrer \&
  Bertschinger}{1991}]{scherrer}
    Scherrer R.~J., Bertschinger E., 1991, ApJ, 381, 349

  \bibitem[\protect\citeauthoryear{Scoccimarro
  et~al.}{2001}]{scoccimarro}
    Scoccimarro R., Sheth R.~K., Hui L., Jain B., 2001, ApJ, 546, 20

  \bibitem[\protect\citeauthoryear{Scranton}{2003}]{scranton}
    Scranton R., 2003, MNRAS, 339, 410

  \bibitem[\protect\citeauthoryear{Seljak}{2000}]{seljak}
    Seljak U., 2000, MNRAS, 318, 203

  \bibitem[\protect\citeauthoryear{Sheth \& Diaferio}{2001}]{diaferio}
    Sheth R.~K., Diaferio A., 2001, MNRAS, 322, 901

  \bibitem[\protect\citeauthoryear{Sheth \& Tormen}{1999}]{sheth_tormen}
    Sheth R.~K., Tormen G., 1999, MNRAS, 308, 119

  \bibitem[\protect\citeauthoryear{Sheth et~al.}{2001}]{sheth_cf}
    Sheth R.~K., Diaferio A., Hui L., Scoccimarro R., 2001, MNRAS,
    326, 423

  \bibitem[\protect\citeauthoryear{Spergel et al.}{2003}]{spergel}
    Spergel D.~N. et~al., 2003, ApJS, 148, 175

  \bibitem[\protect\citeauthoryear{Tegmark et al.}{2002}]{tegmark}
    Tegmark M. et al., 2002, ApJ, 571, 191

  \bibitem[\protect\citeauthoryear{van den Bosch, Yang \& Mo}{van den
  Bosch et al.}{2003}]{vdb}
    van den Bosch F.~C., Yang X., Mo H.~J., 2003, MNRAS, 340, 771

  \bibitem[\protect\citeauthoryear{van den Bosch
  et~al.}{2005}]{vdb_radial}
    van den Bosch F.~C., Yang X., Mo H.~J., Norberg P., 2005, MNRAS,
  356, 1233

  \bibitem[\protect\citeauthoryear{van der Marel et~al.}{2000}]{marel}
    van der Marel R.~P., Magorrian J., Carlberg R.~G., Yee H.~K.~C.,
    Ellingson E., 2000, AJ, 119, 2038

  \bibitem[\protect\citeauthoryear{Wang et~al.}{2004}]{wang}
    Wang Y., Yang X., Mo H.~J., van den Bosch F.~C., Chu Y., 2004,
    MNRAS, 353, 287

  \bibitem[\protect\citeauthoryear{Yang et~al.}{2005a}]{yanga}
    Yang X., Mo H.~J., van den Bosch F.~C., Jing Y.~P., 2005a, MNRAS,
    356, 1293

  \bibitem[\protect\citeauthoryear{Yang et~al.}{2005b}]{yangb}
    Yang X., Mo H.~J., van den Bosch F.~C., Jing Y.~P., 2005b, MNRAS,
    357, 608

  \bibitem[\protect\citeauthoryear{Yang et~al.}{2005c}]{yangc}
    Yang X., Mo H.~J., Jing Y.~P., van den Bosch F.~C., 2005c, MNRAS,
    358, 217

  \bibitem[\protect\citeauthoryear{Zehavi et~al.}{2002}]{zehavi}
    Zehavi I., Blanton M.~R., Frieman J.~A. et~al., 2001, ApJ, 571, 172

\end{thebibliography}
\end{document}